\documentclass[reprint, secnumarabic,amssymb, nobibnotes, aps, prx]{revtex4-1}

\usepackage{color}
\usepackage{epsfig, graphics, graphicx, subfigure, wrapfig, lipsum, longtable,array}
\usepackage{verbatim, import}
\usepackage{dcolumn}% Align table columns on decimal point
\usepackage{bm}% bold math
\usepackage{amsmath, braket}
\usepackage{array}
\usepackage{xr}
\newcolumntype{X}[1]{>{\centering\arraybackslash\hspace{0pt}}p{#1}}
\newcolumntype{M}[1]{ >{\centering\arraybackslash}m{#1}}
\newcommand{\roml}[1]{\lowercase\expandafter{\romannumeral #1\relax}}
\newcommand{\romu}[1]{\uppercase\expandafter{\romannumeral #1\relax}}

\begin{document}

\title{End-to-end Material Thermal Conductivity Prediction through Machine Learning}

\author{Yagyank Srivastava and Ankit Jain}
%\email{a_jain@iitb.ac.in}
\affiliation{Mechanical Engineering Department, IIT Bombay, India}
\date{\today}%

\begin{abstract}
We investigated the accelerated prediction of the thermal conductivity of materials through end-to-end structure-based approaches employing machine learning methods. Due to the non-availability of high-quality thermal conductivity data, we first performed high-throughput calculations based on first principles and the Boltzmann transport equation for 225 materials, effectively more than doubling the size of the existing dataset. We assessed the performance of state-of-the-art machine learning models for thermal conductivity prediction on this expanded dataset and observed that all these models suffered from overfitting. To address this issue, we introduced a novel graph-based neural network model, which demonstrated more consistent and regularized performance across all evaluated datasets. Nevertheless, the best mean absolute percentage error achieved on the test dataset remained in the range of 50-60\%. This suggests that while these models are valuable for expediting material screening, their current accuracy is still limited.
\end{abstract}
\maketitle

\section{Introduction}
%\label{sec_intro}
Thermal conductivity ($\kappa$) is an important material property critical in determining the performance and efficiency of devices in various technological applications such as thermoelectric energy generation, thermal insulation, and memory storage \cite{slack1995, clarke2003, minnich2009a, lindsay2018a}. For many of these applications, low thermal conductivity semiconducting solids are desired, while for others (such as heat dissipation and microprocessors), materials with high $\kappa$ are desired \cite{clarke2003,dames2005,lindsay2018}. For materials used in most of these applications, the thermal transport is dominated by atomic vibrations, i.e., phonons, with room temperature $\kappa$ in the range of $0.1$-$3000$ $\text{W/m-K}$ \cite{ziman1960}. The traditional search for novel low and high $\kappa$ materials is carried out experimentally using a trial-and-error approach. Lately, it has become possible to use ab initio-driven lattice dynamics calculations in conjunction with the Boltzmann transport equation to predict the $\kappa$ of materials \cite{esfarjani2008, esfarjani2011, lindsay2018, mcgaughey2019}. Though these state-of-the-art quantum-mechanical calculations are instrumental in predicting correct $\kappa$ without requiring any fitting parameter \cite{esfarjani2011, lindsay2013b, lindsay2018, jain2020}, the computational cost of these calculations is very high at several hundred to several thousand cpu-hours for each material \cite{esfarjani2008}. Thus, the application of such calculations is limited to simple material systems and computational exploration of new materials is restricted to the simple substitution of one or two atomic species in known material systems \cite{xia2020b,miyazaki2021,  he2021, pal2021}. 

Recently, machine learning (ML) approaches have gained significance for data-driven exploration and discovery of materials.
The ML models can be trained to learn the material properties/behavior from known/training datasets and the trained models can then be used to predict the properties/behavior of new material configurations. Such approaches have already been used successfully to predict the variety of material properties, for instance, ionic conductance \cite{sendek2018}, crystal thermal conductivity \cite{wei2018}, thermoelectric figure of merit \cite{wang2020}, opto-electronic properties \cite{moore2014, pop2010},  mechanical strength \cite{lu2020},  nuclear fuel systems\cite{kautz2019}, and drug discovery \cite{reda2020}. For the particular case of thermal transport, while these approaches are gaining popularity, they are still limited.  
For instance, Pal et al.~\cite{pal2022} employed a scale-invariant ML model to accelerate the search of quaternary chalcogenides with low $\kappa$, Hu et al.~\cite{hu2020} employed ML to minimize coherent heat conduction across aperiodic superlattices, Rodiguez et al.~\cite{rodriguez2023} trained neural network based interatomic forcefield to do bottom-up prediction of $\kappa$ based on intermediate phonon properties such as mean square displacements and bonding/anti-bonding characters, and Visaria and Jain~\cite{visaria2020} employed neural network based auto-encoders to do space transformation to search for material configurations with low- and high-$\kappa$ from the exponentially-large search space of considered superlattices.

For direct end-to-end prediction of $\kappa$ from material crystal structure, the notable contribution is by Zhu et al.~\cite{zhu2021} where log($\kappa$) of diverse material systems is predicted using graph neural network and random forest models and coefficient of determination ($\text{R}^2$) of $0.85$ and $0.87$ are obtained from the two models in a five-fold validation on the dataset consisting of 132 materials. Subsequently, Liu et al.~\cite{liu2022} improved further on this and proposed a transfer learning approach to train a feedforward neural network and obtained an improved $\text{R}^2$ of $0.83$ (compared to $0.67$ with direct learning) on 170 materials in a five-fold cross-validation. 

In both of these studies, however, the obtained performances are reported for five-fold cross-validation and the performances of these models on a different, completely unseen dataset, are not established. This was primarily done due to the limitation on the available datasets with only $\sim$100 entries.
Nonetheless, the evaluation of ML performance on unseen datasets is critical as (\roml{1}) these employed models often have several thousand trainable parameters and as such, are severely prone to over-fitting for datasets of only $\sim$ 100 datapoints, (\roml{2}) while optimizing the model performance, the test dataset is indirectly passed on to ML models via hyper-parameter tuning, and as such, the performance of model on true unseen real-world-like data (completely unseen by ML model) can be different than that on the test data.

With these in mind, in this work, we first carry out a high-throughput, ab-initio driven $\kappa$ calculations to augment the currently available high-quality $\kappa$-dataset by more than two-fold from the current size of 166 to 398 and, next, we develop a graph-based ML model to reduce the problem of over-fitting in $\kappa$ prediction of materials.

\section{Methodology}
\subsection{High-throughput $\kappa$ Calculations}
\label{section_high_throughput}
Whilst the full details regarding the calculation of thermal conductivity using the Boltzmann transport equation approach can be found elsewhere \cite{mcgaughey2019, jain2020}, the thermal conductivity, $\kappa_{\alpha}$, is obtained as \cite{ziman1960, reissland1973, srivastava1990}:
\begin{equation}
 \label{eqn_k}
    \kappa = \sum_i c_{ph, i} v_{\alpha}^2 \tau_i,
\end{equation}
where the summation is over all the phonon modes in the Brillouin zone enumerated by $i\equiv(q,\nu)$, where $q$ and $\nu$ are phonon wavevector and mode index, and $c_{ph,i}$, $v_{\alpha}$, and $\tau_i$ represent phonon specific heat, group velocity ($\alpha$-component), and transport lifetime respectively. The transport lifetimes are obtained by considering phonon-phonon scattering via three-phonon scattering processes.

The harmonic and anharmonic force constants that are required as an input to compute phonon dispersion and phonon scattering rates are obtained from density functional perturbation theory (DFPT) and density functional theory (DFT) calculations as implemented in the openly available quantum mechanical simulation package Quantum Espresso \cite{baroni2001, giannozzi2009}. The planewave-based basis set with norm-conserving Vanderbilt pseudopotentials \cite{schlipf2015} and the planewave kinetic energy cutoff is set at 80 Ry in all calculations. The structure relaxations are performed using primitive unitcells and the electronic Brillouin zone is sampled using a Monkhorst-Pack wavevector grid of size $k_i$ such that $k_i.|a_i^{pri}| \sim 30$ $\text{\AA}$, where $|a_i^{pri}|$ represents the length of primitive unitcell lattice vector $\textbf{a}_i^{pri}$. The electronic total energy is converged to within $10^{-10}$ Ry/atom during self-consistent cycles and the structure relaxations are performed with a force convergence criterion of $10^{-5}$ Ry/$\text{\AA}$.

The harmonic force constants and Born effective charges are obtained using DFPT calculations on primitive unitcells. The DFPT calculations are initially performed on phonon wavevector grids of size $q_i^c$ such that $q_i^c.|a_i^{pri}| \sim 30$ $\text{\AA}$ and are later interpolated to grids of size $q_i^f$ such that $q_i^f.|a_i^{pri}| \sim 100$ $\text{\AA}$ for three-phonon scattering calculations. The anharmonic force constants are obtained from Taylor-series fitting of Hellmann-Feynman forces obtained on 200 thermally populated supercells  (corresponding to a temperature of 300 K) obtained from $N_i$ repetitions of the conventional unitcell such that $N_i.|a_i^{conv}| \sim 15$.  The cubic force constant interaction cutoff is set at $6.5$ for the majority of compounds, though for some compounds with lower symmetries, this value is reduced to $5.0$. 
%{\color{blue}The actual values of simulation parameters are provided in supplementary information for all compounds.}

\begin{figure}
\begin{center}
\epsfbox{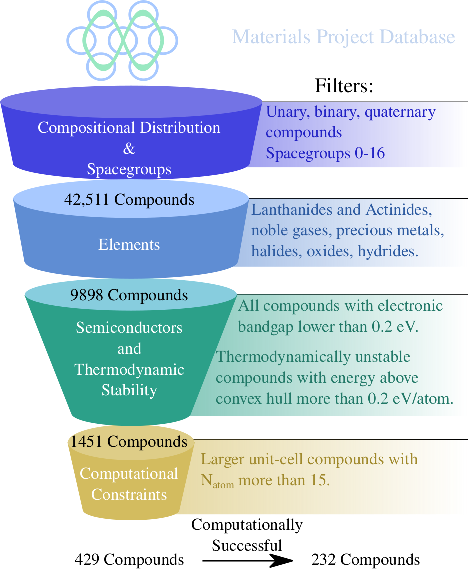}
\end{center}
\caption{The filtering criterion employed to select a pool of stable ternary semiconductors from the Materials Project database \cite{materialsProject} for ab-initio driven high-throughput computation of phonon thermal conductivity.}
\label{fig_screening}
\end{figure}

The material screening process for selecting compounds on which $\kappa$ calculations are carried out is detailed in Fig.~\ref{fig_screening}. Starting with all ternary compounds from the Materials Project \cite{materialsProject} within the orthorhombic, tetragonal, trigonal, hexagonal, and cubic spacegroups, resulting in a total of 42,511 compounds, compounds containing lanthanides and actinides, noble gases, and precious metals (Au, Pt, Pd, Ir, Ru, Re,  Rh, Hg, Hf) are removed, and the maximum atomic number of participating species is restricted to 83. Further, strongly ionic compounds formed by halides, oxides, and hydrides are also removed. Since the focus here is on the phonon thermal transport in semiconductors and insulators, compounds with electronic bandgap lower than $0.2$ eV (as obtained from GGA-based DFT in the Materials Project) are also removed. The materials are further filtered on the basis of thermodynamic stability to have energy above the convex hull (with respect to all reported materials in the Materials Project) less than $0.2$ eV/atom. Finally, considering the $N_{atom}^4$ scaling of computational cost for thermal conductivity calculations \cite{mcgaughey2019}, compounds with more than 15 atoms in the unitcell are also removed. Of the total of 429 filtered compounds, during structure relaxation and dynamical stability calculations, 197 compounds are further dropped due to the presence of imaginary phonon modes and/or non-convergence of self-consistent functional calculations, and the full thermal conductivity calculations are carried out for 232 compounds. 

\subsection{Machine Learning Models}
\label{sec_models}
{\bf Feed-forward Neural Network:}
For the feed-forward neural network (NN), we employed three fully connected layers with a random number of neurons in the first two layers (between 60 to 90 in the first layer and between 20 and 50 in the second layer), and 10 neurons in the final layer with 40 different seedings for initial weights. We generated 360 such models with different combinations of neurons and seedings. We employed Rectified Linear Unit (ReLU) as the activation/non-linear function and used Adam optimizer \cite{kingma2014} with a learning rate of $0.001$ and a batch size of 85. The training is done on 500 epochs with mean absolute error (MAE) loss. For direct feed-forward NN, the weights were randomly initialized from a standard normal distribution while for transfer-learning feed-forward NN, the weights are initialized by pre-training the network on a low-fidelity dataset (discussed in the next Section). This selection of network architecture is similar to that employed in Ref~\cite{liu2022}.

{\bf Graph-based Neural Networks:} For graph-based neural networks,
we employed three different implementations: (a) crystal graph convolution neural network (CGCNN) \cite{xie2017}, (b) materials graph network (MEGNet) \cite{chen2019}, and (c) our own implementation (referred to as GNN). All three of these models are graph-based convolution neural networks and they differ in the implementation of node, edge, and/or state features. For instance, while CGCNN and GNN only have node and edge features, MEGNet also have state features. Similarly, while CGCNN employs one-hot encoding (of length 9) based on interatomic distance for edge features, MEGNet and GNN use only one attribute corresponding to actual value of interatomic spatial distance for edge features. The CGCNN and MEGNet are employed with default settings with MAE training loss and further details on their implementation can be found in Refs \cite{xie2017, chen2019}. 

In GNN, each node represents an atomic site (\textit{i}) in the unitcell with atomic number and position (reduced coordinates) as the node features. The edges represent atomic neighbors within 5 $\text{\AA}$ distance. The atomic numbers are embedded to a vector of length 10 using pytorch embedding \cite{pytorch} which is passed through a non-linear layer to obtain an initial node representation. The node position ($X_{i}$) and neighbor position ($X_{j}$) give the neighbor's distance vector ($X_{j} - X_{i}$), which is passed through a non-linear layer followed by a mean pooling to obtain a consolidated neighbor vector of node (\textit{i}). Finally, node and neighbor vectors are concatenated and passed through a non-linear layer to obtain an updated node representation. This process of obtaining node embedding is repeated three times, and in the third iteration, the lattice parameters ($\alpha, \beta, \gamma, a, b, c$) are concatenated with the consolidated neighbor vector to obtain a final node representation. The final node embeddings are passed through mean and variance pooling followed by a fully connected layer to obtain a prediction for $\kappa$. The GNN model is trained with ReLU activation and  MAE loss using an Adam optimizer with a learning rate of $0.0001$.

{\bf Random Forest and Gaussian Process:} For random forest (RF) and gaussian process (GP), we employed the implementation of Sklearn \cite{sklearn}. For RF, we employed  50 tress and 40 distinct random states. For GP,  we employed a composite kernel by adding three kernels (Radial Basis Function kernel \cite{GPR0,RBFandconst}, Constant Kernel \cite{GPR0,RBFandconst}, and White Kernel \cite{GPR0,White1,White2}) and optimized kernel hyper-parameters with COBYLA method as implemented in Sklearn \cite{sklearn}. In contrast to other ML methods, GP also provides confidence on predicted values which is useful in establishing the uncertainty of predicted $\kappa$.

\subsection{Material Fingerprints} 
For NN, RF, and GP, we generated a material fingerprint using elemental and compound features. For elemental features, we computed five statistical properties (mean, maximum, minimum, etc) of 12 elemental properties (atomic number, electronegativity, atomic weight, etc) of each participating element by employing the Matminer \cite{matminer} ElementProperty interface with "magpie" data source to obtain elemental property vector of length 60. This elemental property vector is concatenated with three compound properties: space group, volume, density, and two Matminer's \cite{matminer} IonProperty (mean and maximum of ionic character) to obtain a final material fingerprint of length 65. For graph-based models, the crystal structures are passed directly to the models, and the material fingerprints are determined internally by the models.

\section{Results}
\subsection{Thermal Conductivity Datasets}
%The datasets employed in this work are visualized using t-distributed stochastic neighbor embedding (t-SNE)  with two components in  Fig.~\ref{fig_metric}.
We considered three $\kappa$-datasets: (\roml{1}) low-fidelity (LF) dataset consisting of 1507 materials with thermal conductivity obtained using a semi-empirical model \cite{liu2022}, (\roml{2}) high-fidelity (HF) dataset consisting of 166 materials whose thermal conductivities are reported in literature from either experimental measurements or from first-principle calculations \cite{liu2022}, and (\roml{3}) high-throughput (HT) dataset consisting of 232 ternary materials whose thermal conductivities are obtained using the full ab-initio calculations with same computational setting for all materials as detailed in Sec.~\ref{section_high_throughput}.  The data contained in LF and HF datasets is isotropic, while the data in HT dataset is direction resolved. As such, to stay consistent with other datasets, we performed directional averaging of $\kappa$ for HT dataset and report all results for direction-averaged mean $\kappa$.
%As such, the actual unique $\kappa$ datapoints in HT dataset are XX though the number of materials are XX. 
The LF and HT datasets have similar $\kappa$ distribution and have most materials with $\kappa$ between 1-20 W/m-K. In comparison, the HF dataset is centered between 10-100 W/m-K and have larger number of compounds with high $\kappa$.
It is worthwhile to emphasize here that even though the size of employed datasets in this study is $\sim$100, these are amongst the largest high-quality datasets available for $\kappa$. On top of a large 3-4 orders of magnitude span of material $\kappa$ values, this scarcity of data also makes ML for $\kappa$ more challenging compared to other material properties.

\begin{figure}
\begin{center}
\epsfbox{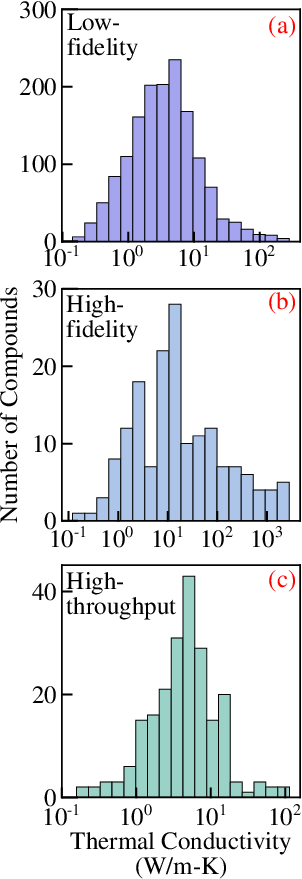}
\end{center}
\caption{The number distribution of $\kappa$ in (a) Low-fidelity \cite{liu2022}, (b) High-fidelity \cite{liu2022}, and (c) High-throughput datasets.}
\label{fig_metric}
\end{figure}

\subsection{Evaluation Metrices}
\label{section_evaluation}
The performance of ML models is generally characterized on hold-out test datasets consisting of 10-30\% of the total data,  using metrics such as mean absolute error (MAE), mean square error (MSE), root mean square error (RMSE), coefficient of determination ($\text{R}^2$), and mean absolute percentage error (MAPE). For properties such as formation energy, electronic bandgap, mixing entropy, etc, the entire range of predicted property is less than an order of magnitude and all of these performace metrices are adequate. For $\kappa$, the range is over four orders of magnitude and, as such, the choice of performance metric is not obvious. 

\begin{figure}
\begin{center}
\epsfbox{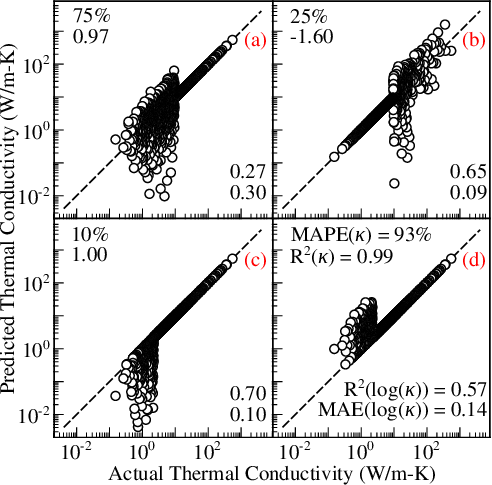}
\end{center}
\caption{The variation of various performance metrics when model predictions are: (a) perfect on all high $\kappa$ and random on low $\kappa$, (b) perfect on all low $\kappa$ and random on high $\kappa$, (c) perfect on all high $\kappa$ and underprediction on low $\kappa$, and (d) perfect on all high $\kappa$ and overprediction on low $\kappa$.The considered datapoints are from LF dataset having the actual distribution of $\kappa$ values.  The $\text{R}^2(\kappa)$ is biased towards high-value datapoints and MAPE($\kappa$) is biased towards overprediction of actual value. }
\label{fig_hypoth}
\end{figure}

To illustrate this on real $\kappa$-dataset, we plot the model performances of four hypothetical models on LF dataset in Fig.~\ref{fig_hypoth}. In Fig.~\ref{fig_hypoth}(a), we considered a model that is able to perfectly predict all high-$\kappa$ points and results in a random prediction for low-$\kappa$ materials and in Fig.~\ref{fig_hypoth}(b), we have a model with perfect prediction for  low-$\kappa$ materials and a random prediction for high-$\kappa$ materials. For Fig.~\ref{fig_hypoth}(a), even though the predictions are random for all low-$\kappa$ materials, the obtained $\text{R}^2$ value is high at $0.97$, while in Fig.~\ref{fig_hypoth}(b), when randomization is applied to high-$\kappa$ points instead,  the $\text{R}^2$ value is negative;  thus demonstrating the bias of $\text{R}^2$ towards large value datapoints. Similarly, in Figs.~\ref{fig_hypoth}(c) and \ref{fig_hypoth}(d), the considered models predict random values for low-$\kappa$ datapoints but they differ in that while model in Fig.~\ref{fig_hypoth}(c) under-predicts, the model in Fig.~\ref{fig_hypoth}(d) over-predicts these low-$\kappa$ datapoints. Since the performance is perfect for all high-$\kappa$ datapoints, the obtained $\text{R}^2$ is close to unity in both cases. However, for MAPE, while the obtained value is low at only 10\% for Fig.~\ref{fig_hypoth}(c), the value is 
 much higher (93\%) for Fig.~\ref{fig_hypoth}(d); thus, indicating the bias of MAPE to under-predict datapoints.

To account for this challenge arising from $\kappa$ spanning over multiple orders of magnitude, the model performance for $\kappa$ prediction is often evaluated on a log-scale where log-$\kappa$ varies in the range $\sim$ -1 to 3. However, this scale is also tricky as the predictions that look good on log-scale are actually severely off due to the exponential nature of this scale. For instance, the predicted log-$\kappa$ for diamond of $3.7$ compared to the experimentally measured value of $3.4$ is off by 100\% even though the difference is less than 10\%  on the log-scale. 

In this work, we report MAE on log $\kappa$ [MAE(log($\kappa$))], MAPE, and $\text{R}^2$ (on $\kappa$ and log($\kappa$)) for all models and we compare model performances based on MAE(log($\kappa$)).

\subsection{Thermal Conductivity Prediction}
\begin{figure*}
\begin{center}
\epsfbox{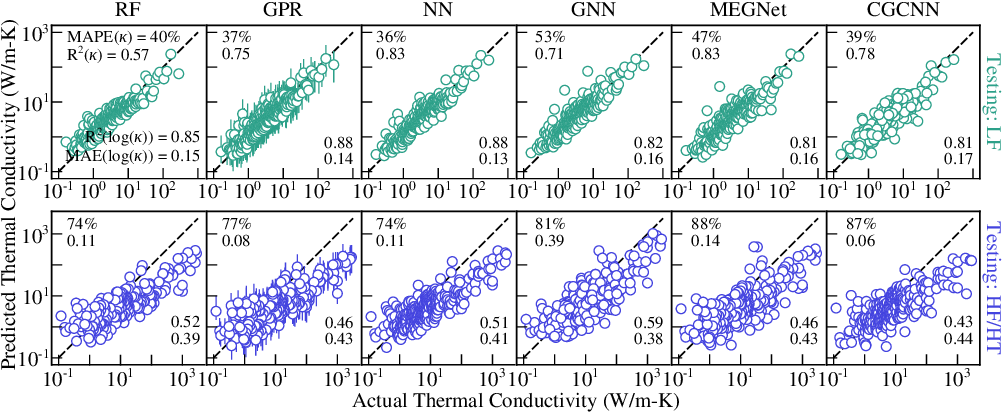}
\end{center}
\caption{The prediction performance of considered ML models on: (top panels) 20\% of holdout compounds from the LF dataset and (bottom panels) HF/HT datasets. All models are trained on 80\% of the LF dataset consisting of the same set of compounds. Without seeing any material from the HF/HT datasets, while all ML models are able to capture the general $\kappa$ trend of these datasets (bottom panels), the best performance is obtained from the GNN model.}
\label{fig_ML_LF_HFHT}
\end{figure*}

We first train the considered ML models on the LF dataset and evaluate their performance for the prediction of $\kappa$ on HF/HT datasets (Fig.~\ref{fig_ML_LF_HFHT}). We find that all ML models can fit the  $\kappa$ variation in the LF dataset correctly with test MAE(log($\kappa$)) of $0.16$ or less. The best train performance is by GPR and NN models, each with test MAE(log($\kappa$)) less than $0.14$. As emphasized above in Sec.~\ref{section_evaluation}, we find that, even though all ML models have test MAE(log($\kappa$)) between $0.14$-$0.16$, the $\text{R}^2$ varied between $0.57$ to $0.83$ depending on the quality of fit of high $\kappa$ datapoints, thus further highlighting the limitations of  $\text{R}^2$ as the performance metric. Note that this limitation is an outcome of orders of magnitude scale of $\kappa$ and when $\text{R}^2$ is evaluated on $\log{\kappa}$, the obtained value is between $0.81$-$0.88$ for all considered models.

When we use these trained models to predict the $\kappa$ of materials from HF/HT datasets, we find that the MAE(log($\kappa$)) of all models is high at $>0.38$. Nevertheless, considering that the LF dataset is derived based on empirical relations and is not a true representative of DFT result/experimental measurements (data in HF/HT datasets), it is impressive to see that all considered models can capture the overall general trend of $\kappa$. The best performance is by the GNN model with MAE(log($\kappa$)) of $0.38$. This same model also resulted in the best generalization between the two datasets as can be estimated based on the ratio of prediction MAE(log($\kappa$)) on HF/HT datasets to test MAE(log($\kappa$)) on the LF dataset. Noticeably, this ratio is only $2.35$ compared to more than $2.6$ for all other models. Furthermore, the GPR and NN models, which resulted in the best performances in testing on the LF dataset, delivered the worst generalization with prediction MAE(log($\kappa$)) of $0.43$ and $0.42$ on the HF/HT dataset; thus suggesting potential over-fitting by these models compared to learning of general trends by the GNN model. It is worthwhile to note here that the number of trainable weights in GNN and NN are similar (between 7000-10000), and, as such, the regularized performance by GNN is due to its specific architecture as presented in Sec.~\ref{sec_models}.

\begin{figure*}
\begin{center}
\epsfbox{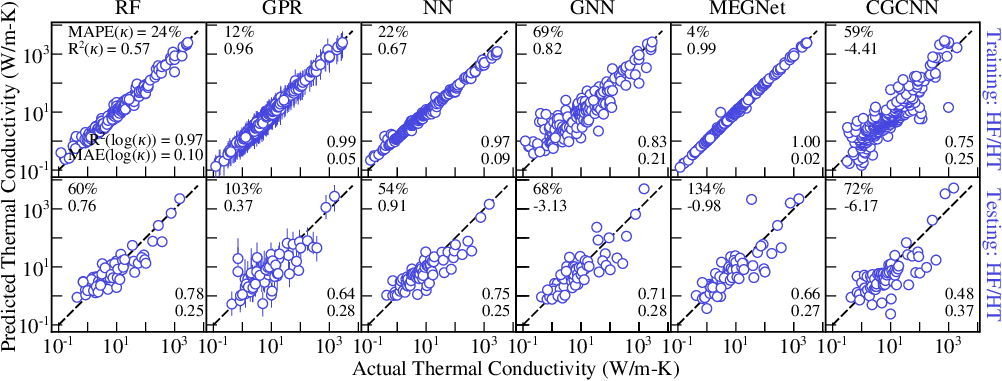}
\end{center}
\caption{The training (top panels) and testing (bottom panels) performance of considered ML models on HF/HT datasets. All models are trained on 80\% of the HF/HT datasets consisting of the same set of compounds. GNN and CGCNN deliver the best-regularized performances, while all other models suffer from over-fitting.}
\label{fig_ML_HFHT}
\end{figure*}

Next, we train the considered ML models directly on HF/HT datasets (total of 398 datapoints) and evaluate model performances on the hold-out dataset (Fig.~\ref{fig_ML_HFHT}). We generated the train-test split by sorting the materials by their $\kappa$ and by placing every $5^{th}$ entry from this sorted list in the test dataset (80:20 split) for all models. 
%{\color{red} Further, if any compound in test set had elements which are unseen to train set, a swap was done with compound of similar $\kappa$ between train and test set.}
During training, we find that the MAE(log($\kappa$)) for RF, GPR, NN, and MegNet models reduced compared to their corresponding test values for the LF dataset. For instance, for NN, the MAE(log($\kappa$)) reduced from $0.14$ for testing on the LF dataset to $0.04$ for training on HF/HT datasets. This reduction is understandable as the number of datapoints used in HF/HT training is only 398 compared to 1507 on the LF dataset. Surprisingly, for GNN and CGCNN, this is not true, and the MAE(log($\kappa$)) is higher for training on HF/HT datasets than for testing on the LF dataset. Both GNN and CGCNN resulted in MAE(log($\kappa$)) higher than $0.21$ during training on the HF/HT compared to less than $0.10$ by all other models. This large  MAE(log($\kappa$)) from GNN and CGCNN seems to suggest the inferior performance of these models compared to other models, but when we employed the trained models to predict $\kappa$ of hold-out test dataset (75 datapoints), we found all models resulted in test MAE(log($\kappa$)) $>0.25$, thus suggesting over-fitting by all models. Compared to other models, for GNN and CGCNN, the over-fitting is less pronounced. For GNN and CGCNN, the test errors are only 36\%  and 46\% larger than the corresponding train errors, but for RF, GPR, NN, and MeGNet, the test errors are 161\%, 571\%, 166\%, and 1411\% larger than the train errors. For NN, we also included regularization via the Dropout method \cite{srivastava2014} (after first hidden layer) with 20\% and 50\% dropout rates, but in both cases, we found that the obtained test error was $>160\%$ larger than the training error. 

\begin{comment}
\begin{figure}
\begin{center}
\epsfbox{Figures/6.eps}
\end{center}
\caption{Actual vs predicted plot of training (blue) and testing (green) on HT and HF data combined, after inclusion of dropout regularization with probability of a), b) 0.2 and c),d) 0.5 in model architecture of NN.}
\label{fig_NN_Dropout}
\end{figure}
\end{comment}

\begin{figure}
\begin{center}
\epsfbox{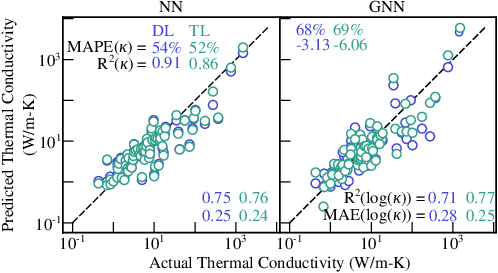}
\end{center}
\caption{The effect of transfer learning (TL) on the prediction performance of (left) NN model and (right) GNN model. In direct learning (DL), the models are directly trained on the HF/HT datasets while in TL, the models are first pre-trained on the LF dataset and are later re-trained on the HF/HT datasets. 
The reported performances are on the 20\% of the hold-out dataset.}
\label{fig_TLDL}
\end{figure}

Motivated by this impressive performance of GNN, along with its associated regularization, we explore the effect of transfer learning on GNN by pre-training it first on the LF dataset, followed by training on the HF/HT datasets (Fig.~\ref{fig_TLDL}). For reference, we also included the NN model. As suggested by Liu et al.~\cite{liu2022} for NN, we find that the prediction performance of GNN also improves with transfer learning. The test MAE(log($\kappa$)) from GNN is reduced to $0.25$ with transfer learning, which is similar to that from the NN model. 

For materials exploration, it is imperative to have a confidence interval on the predicted value of $\kappa$ from ML models. For the GPR model, the confidence intervals are analytically obtained from the model itself depending on the kernel similarity between train and test datapoints. However, as shown earlier, for $\kappa$ prediction, GPR results in a severe over-fitting and the predicted $\kappa$ from GPR model is out of one standard deviation bounds for a large number of test datapoints in Fig.~\ref{fig_ML_HFHT}. As an alternative, we pick the predicted $\kappa$ variation between different ML models to measure confidence bounds.

\begin{figure}
\begin{center}
\epsfbox{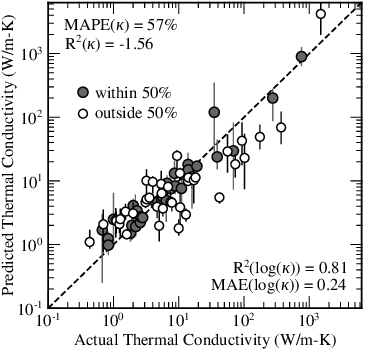}
\end{center}
\caption{The prediction performance of the combined model on 20\% of holdout HF/HT datasets. The predicted values are taken as mean of RF, NN, GNN, and CGCNN models and the plotted lower/upper bounds are minimum/maximum predictions from these models.
}
\label{fig_confidence}
\end{figure}

In Fig.~\ref{fig_confidence}, the predicted $\kappa$ for test materials from HF/HT datasets is re-plotted by combining predictions from RF, NN, GNN, and CGCNN models (GPR and MEGNet are omitted as both of these models result in severe over-fitting with test error being more than $500$\% larger than the train error). The lower/upper bounds are obtained by taking the minimum/maximum over all models, and the predicted value is obtained by taking the average of all models. Further, the materials for which true $\kappa$ falls outside the lower/upper bounds are highlighted using empty markers in Fig.~\ref{fig_confidence}. 

With this combined model, we find that the prediction performance is further improved and MAE(log($\kappa$)) is reduced to $0.23$. Further, the actual $\kappa$  falls within this combined model's lower/upper bounds for around 50\% of the 75 materials tested in Fig.~\ref{fig_confidence}.

\begin{figure}
\begin{center}
\epsfbox{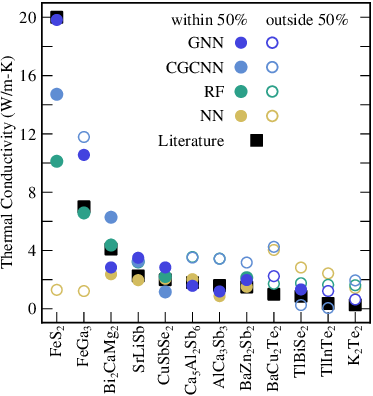}
\end{center}
\caption{The prediction performance of considered ML models on 12 completely unseen materials which are recently reported in the literature \cite{lit1,lit2,lit3,lit4,lit5,lit6,lit7,lit8,lit9,lit10}. The predictions which are within 50\% of the literature-reported value are indicated using solid circles and those which are outside of 50\% are denoted using open circles.}
\label{fig_literature}
\end{figure}

Finally, we also tested the performance of trained ML models on some of the recently reported materials from the literature \cite{lit1,lit2,lit3,lit4,lit5,lit6,lit7,lit8,lit9,lit10} after ensuring that these materials are not a part of any of the databases employed in model training/testing. We report these performances in Fig.~\ref{fig_literature}. The predictions that fall within 50\% of the literature-reported values are indicated using filled circles, while those falling outside of 50\% are indicated using open circles in Fig.~\ref{fig_literature}. We find that for extremely low $\kappa$, while predicted trends are right, all model predictions are off by more than 50\% from the literature-reported values. This is, however, understandable as the lowest $\kappa$ in our training dataset is $0.47$ W/m-K. Surprisingly, the NN model seems to predict values between 1-2 W/m-K for all materials and results in a severe under-prediction for intermediate $\kappa$ materials (with $\kappa>4$ W/m-K in Fig.~\ref{fig_literature}). For materials with $\kappa > 0.47$ W/m-K, the GNN predictions are off by more than 50\% only for two material as compared to five by RF and CGCNN; thus, further highlighting the superior performance of GNN compared to all other ML models employed in this work.
 
\section{Discussion}
The application of ML methods for thermal transport can be broadly classified into three categories. In the first kind of application of ML for thermal properties, instead of direct prediction of $\kappa$/thermal properties, the interatomic forcefields are trained on DFT forces/energies and these trained forcefields are used in conjunction with approaches such as molecular dynamics simulations, lattice dynamics calculations, etc to predict the thermal properties. These DFT-trained ML forcefields offer the advantage of capturing the right thermal transport physics (depending on the approach) and other than the employed approach, the accuracy of results depends entirely on the accuracy of the trained forcefield (which can now approach the same accuracy as that in DFT \cite{MLFF}). Further, this DFT-trained ML forcefield is generic and it could be employed to predict other material properties of a given system. However, this trained forcefield is not transferable and specific to a given material(s) (for which it is trained). Further, one needs to look into the computational cost of data generation for this forcefield training compared to the speedup obtained from employing this forcefield for studying thermal properties. 

Alternatively, ML can be used to explore the chemical/configuration space of a given material system to identify hidden trends and later employ the knowledge gained in this exploration to identify configurations with desired properties. The objective in this approach is to learn the property trends over a large configuration space (potentially exponentially large), which is otherwise difficult to obtain. Examples of such applications include the study by Hu et al.~\cite{hu2020} to minimize coherent conduction across aperiodic interfaces and the study by Visaria and Jain \cite{visaria2020} to explore high/low thermal $\kappa$ materials from exponentially large search of graphene composed of light/heavy carbon atoms.

Finally, the ML can also be used to directly predict $\kappa$ from the material structure or some other intermediate thermal/material property (such as heat capacity, speed of sound, etc). While the direct prediction of $\kappa$ from the structure is a holy grail for the discovery of low/high $\kappa$ solids for various applications, it is currently limited due to the availability of only a sparse amount of data, which is insufficient in exploring the entire chemical space of inorganic materials. While approaches such as transfer learning are useful in such scenarios, our study suggests that the application of such approaches for $\kappa$ have, so far, resulted in the best MAE(log($\kappa$)) of $0.23$ and $\text{R}^2 (\log(\kappa))$ of $0.81$ with MAPE of around 55\%.

\section{Conclusions}
In summary, we explored the possibility of end-to-end structure to thermal conductivity prediction of materials using various machine learning approaches. Due to the scarcity of available thermal conductivity data, we first carried out high-throughput first-principles-based Boltzmann transport equation-driven prediction of the thermal conductivity to augment the currently available dataset from a current size of 166 to 398 materials. 
We tested various performance evaluation metrices and found that all linear-scale evaluation metrices are biased either toward under-prediction or towards high-value datapoints. We find that literature-reported best machine learning models are prone to over-fitting due to scarcity of data and we proposed a new graph based network architecture capable of delivering more regularized and consistent performance across considered smaller size thermal conductivity datasets. Transfer learning by first training the models on larger size, lower accuracy dataset improves model performance but the best-obtained performance from considered models is limited to mean absolute percentage error of $\sim$60\%, even on an augmented dataset (which has more than twice as many entries as the largest reported datasets available in the literature) with the newly proposed graph model (which delivers superior performance compared to all other models); thus suggesting that while machine learning can be used for initial screening of materials, the currently obtainable accuracy for end-to-end thermal conductivity prediction is limited to 50-60\%.

\section{Acknowledgement}
The authors acknowledge the financial support from the National Supercomputing Mission, Government of India (Grant Number: DST/NSM/R\&D-HPC-Applications/2021/10) and Core Research Grant, Science \& Engineering Research Board, India (Grant Number: CRG/2021/000010).  The calculations are carried out on the SpaceTime-II supercomputing facility of IIT Bombay and the PARAM Sanganak supercomputing facility of IIT Kanpur. The authors acknowledge useful discussions with Shravan Godse at IIT Bombay.

\section{Data Availability}
The data that support the findings of this study are available from the corresponding author upon reasonable request.

%\bibliography{references, references_ML}

\begin{thebibliography}{58}%
\makeatletter
\providecommand \@ifxundefined [1]{%
 \@ifx{#1\undefined}
}%
\providecommand \@ifnum [1]{%
 \ifnum #1\expandafter \@firstoftwo
 \else \expandafter \@secondoftwo
 \fi
}%
\providecommand \@ifx [1]{%
 \ifx #1\expandafter \@firstoftwo
 \else \expandafter \@secondoftwo
 \fi
}%
\providecommand \natexlab [1]{#1}%
\providecommand \enquote  [1]{``#1''}%
\providecommand \bibnamefont  [1]{#1}%
\providecommand \bibfnamefont [1]{#1}%
\providecommand \citenamefont [1]{#1}%
\providecommand \href@noop [0]{\@secondoftwo}%
\providecommand \href [0]{\begingroup \@sanitize@url \@href}%
\providecommand \@href[1]{\@@startlink{#1}\@@href}%
\providecommand \@@href[1]{\endgroup#1\@@endlink}%
\providecommand \@sanitize@url [0]{\catcode `\\12\catcode `\$12\catcode
  `\&12\catcode `\#12\catcode `\^12\catcode `\_12\catcode `\%12\relax}%
\providecommand \@@startlink[1]{}%
\providecommand \@@endlink[0]{}%
\providecommand \url  [0]{\begingroup\@sanitize@url \@url }%
\providecommand \@url [1]{\endgroup\@href {#1}{\urlprefix }}%
\providecommand \urlprefix  [0]{URL }%
\providecommand \Eprint [0]{\href }%
\providecommand \doibase [0]{http://dx.doi.org/}%
\providecommand \selectlanguage [0]{\@gobble}%
\providecommand \bibinfo  [0]{\@secondoftwo}%
\providecommand \bibfield  [0]{\@secondoftwo}%
\providecommand \translation [1]{[#1]}%
\providecommand \BibitemOpen [0]{}%
\providecommand \bibitemStop [0]{}%
\providecommand \bibitemNoStop [0]{.\EOS\space}%
\providecommand \EOS [0]{\spacefactor3000\relax}%
\providecommand \BibitemShut  [1]{\csname bibitem#1\endcsname}%
\let\auto@bib@innerbib\@empty
%</preamble>
\bibitem [{\citenamefont {Slack}(1995)}]{slack1995}%
  \BibitemOpen
  \bibfield  {author} {\bibinfo {author} {\bibfnamefont {G.~A.}\ \bibnamefont
  {Slack}},\ }in\ \href@noop {} {\emph {\bibinfo {booktitle} {CRC Handbook of
  Thermoelectrics}}},\ \bibinfo {editor} {edited by\ \bibinfo {editor}
  {\bibfnamefont {D.~M.}\ \bibnamefont {Rowe}}}\ (\bibinfo  {publisher} {CRC,
  Boca Raton},\ \bibinfo {year} {1995})\ pp.\ \bibinfo {pages}
  {407--440}\BibitemShut {NoStop}%
\bibitem [{\citenamefont {Clarke}(2003)}]{clarke2003}%
  \BibitemOpen
  \bibfield  {author} {\bibinfo {author} {\bibfnamefont {D.~R.}\ \bibnamefont
  {Clarke}},\ }\href@noop {} {\bibfield  {journal} {\bibinfo  {journal}
  {Surface and Coatings Technology}\ }\textbf {\bibinfo {volume} {163}},\
  \bibinfo {pages} {67} (\bibinfo {year} {2003})}\BibitemShut {NoStop}%
\bibitem [{\citenamefont {Minnich}\ \emph {et~al.}(2009)\citenamefont
  {Minnich}, \citenamefont {Dresselhaus}, \citenamefont {Ren},\ and\
  \citenamefont {Chen}}]{minnich2009a}%
  \BibitemOpen
  \bibfield  {author} {\bibinfo {author} {\bibfnamefont {A.~J.}\ \bibnamefont
  {Minnich}}, \bibinfo {author} {\bibfnamefont {M.~S.}\ \bibnamefont
  {Dresselhaus}}, \bibinfo {author} {\bibfnamefont {F.}~\bibnamefont {Ren}}, \
  and\ \bibinfo {author} {\bibfnamefont {G.}~\bibnamefont {Chen}},\ }\href@noop
  {} {\bibfield  {journal} {\bibinfo  {journal} {Energy and Environmental
  Sciences}\ }\textbf {\bibinfo {volume} {2}},\ \bibinfo {pages} {466}
  (\bibinfo {year} {2009})}\BibitemShut {NoStop}%
\bibitem [{\citenamefont {Lindsay}\ and\ \citenamefont
  {Polanco}(2018)}]{lindsay2018a}%
  \BibitemOpen
  \bibfield  {author} {\bibinfo {author} {\bibfnamefont {L.}~\bibnamefont
  {Lindsay}}\ and\ \bibinfo {author} {\bibfnamefont {C.~A.}\ \bibnamefont
  {Polanco}},\ }\href {\doibase 10.1007/978-3-319-50257-1_10-1} {\emph
  {\bibinfo {title} {Handbook of Materials Modeling}}}\ (\bibinfo {year}
  {2018})\ pp.\ \bibinfo {pages} {1--31}\BibitemShut {NoStop}%
\bibitem [{\citenamefont {Dames}\ and\ \citenamefont {Chen}(2005)}]{dames2005}%
  \BibitemOpen
  \bibfield  {author} {\bibinfo {author} {\bibfnamefont {C.}~\bibnamefont
  {Dames}}\ and\ \bibinfo {author} {\bibfnamefont {G.}~\bibnamefont {Chen}},\
  }\enquote {\bibinfo {title} {Thermoelectrics handbook: macro to nano},}\ \
  (\bibinfo  {publisher} {CRC press},\ \bibinfo {year} {2005})\ Chap.~\bibinfo
  {chapter} {42}, pp.\ \bibinfo {pages} {42--1--42--11}\BibitemShut {NoStop}%
\bibitem [{\citenamefont {Lindsay}\ \emph {et~al.}(2018)\citenamefont
  {Lindsay}, \citenamefont {Hua}, \citenamefont {Ruan},\ and\ \citenamefont
  {Lee}}]{lindsay2018}%
  \BibitemOpen
  \bibfield  {author} {\bibinfo {author} {\bibfnamefont {L.}~\bibnamefont
  {Lindsay}}, \bibinfo {author} {\bibfnamefont {C.}~\bibnamefont {Hua}},
  \bibinfo {author} {\bibfnamefont {X.}~\bibnamefont {Ruan}}, \ and\ \bibinfo
  {author} {\bibfnamefont {S.}~\bibnamefont {Lee}},\ }\href@noop {} {\bibfield
  {journal} {\bibinfo  {journal} {Materials Today Physics}\ }\textbf {\bibinfo
  {volume} {7}},\ \bibinfo {pages} {106} (\bibinfo {year} {2018})}\BibitemShut
  {NoStop}%
\bibitem [{\citenamefont {Ziman}(1960)}]{ziman1960}%
  \BibitemOpen
  \bibfield  {author} {\bibinfo {author} {\bibfnamefont {J.~M.}\ \bibnamefont
  {Ziman}},\ }\href@noop {} {\emph {\bibinfo {title} {Electrons and Phonons}}}\
  (\bibinfo  {publisher} {Oxford University Press, Clarendon, Oxford},\
  \bibinfo {year} {1960})\BibitemShut {NoStop}%
\bibitem [{\citenamefont {Esfarjani}\ and\ \citenamefont
  {Stokes}(2008)}]{esfarjani2008}%
  \BibitemOpen
  \bibfield  {author} {\bibinfo {author} {\bibfnamefont {K.}~\bibnamefont
  {Esfarjani}}\ and\ \bibinfo {author} {\bibfnamefont {H.~T.}\ \bibnamefont
  {Stokes}},\ }\href@noop {} {\bibfield  {journal} {\bibinfo  {journal}
  {Physical Review B}\ }\textbf {\bibinfo {volume} {77}},\ \bibinfo {pages}
  {144112} (\bibinfo {year} {2008})}\BibitemShut {NoStop}%
\bibitem [{\citenamefont {Esfarjani}\ \emph {et~al.}(2011)\citenamefont
  {Esfarjani}, \citenamefont {Chen},\ and\ \citenamefont
  {Stokes}}]{esfarjani2011}%
  \BibitemOpen
  \bibfield  {author} {\bibinfo {author} {\bibfnamefont {K.}~\bibnamefont
  {Esfarjani}}, \bibinfo {author} {\bibfnamefont {G.}~\bibnamefont {Chen}}, \
  and\ \bibinfo {author} {\bibfnamefont {H.~T.}\ \bibnamefont {Stokes}},\
  }\href@noop {} {\bibfield  {journal} {\bibinfo  {journal} {Physical Review
  B}\ }\textbf {\bibinfo {volume} {84}},\ \bibinfo {pages} {085204} (\bibinfo
  {year} {2011})}\BibitemShut {NoStop}%
\bibitem [{\citenamefont {McGaughey}\ \emph {et~al.}(2019)\citenamefont
  {McGaughey}, \citenamefont {Jain}, \citenamefont {Kim},\ and\ \citenamefont
  {Fu}}]{mcgaughey2019}%
  \BibitemOpen
  \bibfield  {author} {\bibinfo {author} {\bibfnamefont {A.~J.}\ \bibnamefont
  {McGaughey}}, \bibinfo {author} {\bibfnamefont {A.}~\bibnamefont {Jain}},
  \bibinfo {author} {\bibfnamefont {H.~Y.}\ \bibnamefont {Kim}}, \ and\
  \bibinfo {author} {\bibfnamefont {B.}~\bibnamefont {Fu}},\ }\href {\doibase
  10.1063/1.5064602} {\bibfield  {journal} {\bibinfo  {journal} {Journal of
  Applied Physics}\ }\textbf {\bibinfo {volume} {125}},\ \bibinfo {pages}
  {11101} (\bibinfo {year} {2019})}\BibitemShut {NoStop}%
\bibitem [{\citenamefont {Lindsay}\ \emph {et~al.}(2013)\citenamefont
  {Lindsay}, \citenamefont {Broido},\ and\ \citenamefont
  {Reinecke}}]{lindsay2013b}%
  \BibitemOpen
  \bibfield  {author} {\bibinfo {author} {\bibfnamefont {L.}~\bibnamefont
  {Lindsay}}, \bibinfo {author} {\bibfnamefont {D.~A.}\ \bibnamefont {Broido}},
  \ and\ \bibinfo {author} {\bibfnamefont {T.~L.}\ \bibnamefont {Reinecke}},\
  }\href {\doibase 10.1103/PhysRevB.87.165201} {\bibfield  {journal} {\bibinfo
  {journal} {Phys. Rev. B}\ }\textbf {\bibinfo {volume} {87}},\ \bibinfo
  {pages} {165201} (\bibinfo {year} {2013})}\BibitemShut {NoStop}%
\bibitem [{\citenamefont {Jain}(2020)}]{jain2020}%
  \BibitemOpen
  \bibfield  {author} {\bibinfo {author} {\bibfnamefont {A.}~\bibnamefont
  {Jain}},\ }\href {\doibase 10.1103/physrevb.102.201201} {\bibfield  {journal}
  {\bibinfo  {journal} {Physical Review B}\ }\textbf {\bibinfo {volume} {102}}
  (\bibinfo {year} {2020}),\ 10.1103/physrevb.102.201201}\BibitemShut {NoStop}%
\bibitem [{\citenamefont {Xia}\ \emph {et~al.}(2020)\citenamefont {Xia},
  \citenamefont {Hegde}, \citenamefont {Pal}, \citenamefont {Hua},
  \citenamefont {Gaines}, \citenamefont {Patel}, \citenamefont {He},
  \citenamefont {Aykol},\ and\ \citenamefont {Wolverton}}]{xia2020b}%
  \BibitemOpen
  \bibfield  {author} {\bibinfo {author} {\bibfnamefont {Y.}~\bibnamefont
  {Xia}}, \bibinfo {author} {\bibfnamefont {V.~I.}\ \bibnamefont {Hegde}},
  \bibinfo {author} {\bibfnamefont {K.}~\bibnamefont {Pal}}, \bibinfo {author}
  {\bibfnamefont {X.}~\bibnamefont {Hua}}, \bibinfo {author} {\bibfnamefont
  {D.}~\bibnamefont {Gaines}}, \bibinfo {author} {\bibfnamefont
  {S.}~\bibnamefont {Patel}}, \bibinfo {author} {\bibfnamefont
  {J.}~\bibnamefont {He}}, \bibinfo {author} {\bibfnamefont {M.}~\bibnamefont
  {Aykol}}, \ and\ \bibinfo {author} {\bibfnamefont {C.}~\bibnamefont
  {Wolverton}},\ }\href@noop {} {\bibfield  {journal} {\bibinfo  {journal}
  {Physical Review X}\ }\textbf {\bibinfo {volume} {10}},\ \bibinfo {pages}
  {041029} (\bibinfo {year} {2020})}\BibitemShut {NoStop}%
\bibitem [{\citenamefont {Miyazaki}\ \emph {et~al.}(2021)\citenamefont
  {Miyazaki}, \citenamefont {Tamura}, \citenamefont {Mikami}, \citenamefont
  {Watanabe}, \citenamefont {Ide}, \citenamefont {Ozkendir},\ and\
  \citenamefont {Nishino}}]{miyazaki2021}%
  \BibitemOpen
  \bibfield  {author} {\bibinfo {author} {\bibfnamefont {H.}~\bibnamefont
  {Miyazaki}}, \bibinfo {author} {\bibfnamefont {T.}~\bibnamefont {Tamura}},
  \bibinfo {author} {\bibfnamefont {M.}~\bibnamefont {Mikami}}, \bibinfo
  {author} {\bibfnamefont {K.}~\bibnamefont {Watanabe}}, \bibinfo {author}
  {\bibfnamefont {N.}~\bibnamefont {Ide}}, \bibinfo {author} {\bibfnamefont
  {O.~M.}\ \bibnamefont {Ozkendir}}, \ and\ \bibinfo {author} {\bibfnamefont
  {Y.}~\bibnamefont {Nishino}},\ }\href@noop {} {\bibfield  {journal} {\bibinfo
   {journal} {Scientific reports}\ }\textbf {\bibinfo {volume} {11}},\ \bibinfo
  {pages} {1} (\bibinfo {year} {2021})}\BibitemShut {NoStop}%
\bibitem [{\citenamefont {He}\ \emph {et~al.}()\citenamefont {He},
  \citenamefont {Xia}, \citenamefont {Lin}, \citenamefont {Pal}, \citenamefont
  {Zhu}, \citenamefont {Kanatzidis},\ and\ \citenamefont {Wolverton}}]{he2021}%
  \BibitemOpen
  \bibfield  {author} {\bibinfo {author} {\bibfnamefont {J.}~\bibnamefont
  {He}}, \bibinfo {author} {\bibfnamefont {Y.}~\bibnamefont {Xia}}, \bibinfo
  {author} {\bibfnamefont {W.}~\bibnamefont {Lin}}, \bibinfo {author}
  {\bibfnamefont {K.}~\bibnamefont {Pal}}, \bibinfo {author} {\bibfnamefont
  {Y.}~\bibnamefont {Zhu}}, \bibinfo {author} {\bibfnamefont {M.~G.}\
  \bibnamefont {Kanatzidis}}, \ and\ \bibinfo {author} {\bibfnamefont
  {C.}~\bibnamefont {Wolverton}},\ }\href@noop {} {\bibinfo  {journal}
  {Advanced Functional Materials}\ ,\ \bibinfo {pages} {2108532}}\BibitemShut
  {NoStop}%
\bibitem [{\citenamefont {Pal}\ \emph {et~al.}(2021)\citenamefont {Pal},
  \citenamefont {Xia}, \citenamefont {Shen}, \citenamefont {He}, \citenamefont
  {Luo}, \citenamefont {Kanatzidis},\ and\ \citenamefont
  {Wolverton}}]{pal2021}%
  \BibitemOpen
\bibfield  {journal} {  }\bibfield  {author} {\bibinfo {author} {\bibfnamefont
  {K.}~\bibnamefont {Pal}}, \bibinfo {author} {\bibfnamefont {Y.}~\bibnamefont
  {Xia}}, \bibinfo {author} {\bibfnamefont {J.}~\bibnamefont {Shen}}, \bibinfo
  {author} {\bibfnamefont {J.}~\bibnamefont {He}}, \bibinfo {author}
  {\bibfnamefont {Y.}~\bibnamefont {Luo}}, \bibinfo {author} {\bibfnamefont
  {M.~G.}\ \bibnamefont {Kanatzidis}}, \ and\ \bibinfo {author} {\bibfnamefont
  {C.}~\bibnamefont {Wolverton}},\ }\href@noop {} {\bibfield  {journal}
  {\bibinfo  {journal} {npj Computational Materials}\ }\textbf {\bibinfo
  {volume} {7}},\ \bibinfo {pages} {1} (\bibinfo {year} {2021})}\BibitemShut
  {NoStop}%
\bibitem [{\citenamefont {Sendek}\ \emph {et~al.}(2018)\citenamefont {Sendek},
  \citenamefont {Cubuk}, \citenamefont {Antoniuk}, \citenamefont {Cheon},
  \citenamefont {Cui},\ and\ \citenamefont {Reed}}]{sendek2018}%
  \BibitemOpen
  \bibfield  {author} {\bibinfo {author} {\bibfnamefont {A.~D.}\ \bibnamefont
  {Sendek}}, \bibinfo {author} {\bibfnamefont {E.~D.}\ \bibnamefont {Cubuk}},
  \bibinfo {author} {\bibfnamefont {E.~R.}\ \bibnamefont {Antoniuk}}, \bibinfo
  {author} {\bibfnamefont {G.}~\bibnamefont {Cheon}}, \bibinfo {author}
  {\bibfnamefont {Y.}~\bibnamefont {Cui}}, \ and\ \bibinfo {author}
  {\bibfnamefont {E.~J.}\ \bibnamefont {Reed}},\ }\href@noop {} {\bibfield
  {journal} {\bibinfo  {journal} {arXiv preprint arXiv:1808.02470}\ } (\bibinfo
  {year} {2018})}\BibitemShut {NoStop}%
\bibitem [{\citenamefont {Wei}\ \emph {et~al.}(2018)\citenamefont {Wei},
  \citenamefont {Zhao}, \citenamefont {Rong},\ and\ \citenamefont
  {Bao}}]{wei2018}%
  \BibitemOpen
  \bibfield  {author} {\bibinfo {author} {\bibfnamefont {H.}~\bibnamefont
  {Wei}}, \bibinfo {author} {\bibfnamefont {S.}~\bibnamefont {Zhao}}, \bibinfo
  {author} {\bibfnamefont {Q.}~\bibnamefont {Rong}}, \ and\ \bibinfo {author}
  {\bibfnamefont {H.}~\bibnamefont {Bao}},\ }\href@noop {} {\bibfield
  {journal} {\bibinfo  {journal} {International Journal of Heat and Mass
  Transfer}\ }\textbf {\bibinfo {volume} {127}},\ \bibinfo {pages} {908}
  (\bibinfo {year} {2018})}\BibitemShut {NoStop}%
\bibitem [{\citenamefont {Wang}\ \emph {et~al.}(2020)\citenamefont {Wang},
  \citenamefont {Zhang}, \citenamefont {Snoussi},\ and\ \citenamefont
  {Zhang}}]{wang2020}%
  \BibitemOpen
  \bibfield  {author} {\bibinfo {author} {\bibfnamefont {T.}~\bibnamefont
  {Wang}}, \bibinfo {author} {\bibfnamefont {C.}~\bibnamefont {Zhang}},
  \bibinfo {author} {\bibfnamefont {H.}~\bibnamefont {Snoussi}}, \ and\
  \bibinfo {author} {\bibfnamefont {G.}~\bibnamefont {Zhang}},\ }\href@noop {}
  {\bibfield  {journal} {\bibinfo  {journal} {Advanced Functional Materials}\
  }\textbf {\bibinfo {volume} {30}},\ \bibinfo {pages} {1906041} (\bibinfo
  {year} {2020})}\BibitemShut {NoStop}%
\bibitem [{\citenamefont {Moore}\ and\ \citenamefont {Shi}(2014)}]{moore2014}%
  \BibitemOpen
  \bibfield  {author} {\bibinfo {author} {\bibfnamefont {A.~L.}\ \bibnamefont
  {Moore}}\ and\ \bibinfo {author} {\bibfnamefont {L.}~\bibnamefont {Shi}},\
  }\href@noop {} {\bibfield  {journal} {\bibinfo  {journal} {Materials today}\
  }\textbf {\bibinfo {volume} {17}},\ \bibinfo {pages} {163} (\bibinfo {year}
  {2014})}\BibitemShut {NoStop}%
\bibitem [{\citenamefont {Pop}(2010)}]{pop2010}%
  \BibitemOpen
  \bibfield  {author} {\bibinfo {author} {\bibfnamefont {E.}~\bibnamefont
  {Pop}},\ }\href@noop {} {\bibfield  {journal} {\bibinfo  {journal} {Nano
  Research}\ }\textbf {\bibinfo {volume} {3}},\ \bibinfo {pages} {147}
  (\bibinfo {year} {2010})}\BibitemShut {NoStop}%
\bibitem [{\citenamefont {Lu}\ \emph {et~al.}(2020)\citenamefont {Lu},
  \citenamefont {Dao}, \citenamefont {Kumar}, \citenamefont {Ramamurty},
  \citenamefont {Karniadakis},\ and\ \citenamefont {Suresh}}]{lu2020}%
  \BibitemOpen
  \bibfield  {author} {\bibinfo {author} {\bibfnamefont {L.}~\bibnamefont
  {Lu}}, \bibinfo {author} {\bibfnamefont {M.}~\bibnamefont {Dao}}, \bibinfo
  {author} {\bibfnamefont {P.}~\bibnamefont {Kumar}}, \bibinfo {author}
  {\bibfnamefont {U.}~\bibnamefont {Ramamurty}}, \bibinfo {author}
  {\bibfnamefont {G.~E.}\ \bibnamefont {Karniadakis}}, \ and\ \bibinfo {author}
  {\bibfnamefont {S.}~\bibnamefont {Suresh}},\ }\href@noop {} {\bibfield
  {journal} {\bibinfo  {journal} {Proceedings of the National Academy of
  Sciences}\ }\textbf {\bibinfo {volume} {117}},\ \bibinfo {pages} {7052}
  (\bibinfo {year} {2020})}\BibitemShut {NoStop}%
\bibitem [{\citenamefont {Kautz}\ \emph {et~al.}(2019)\citenamefont {Kautz},
  \citenamefont {Hagen}, \citenamefont {Johns},\ and\ \citenamefont
  {Burkes}}]{kautz2019}%
  \BibitemOpen
  \bibfield  {author} {\bibinfo {author} {\bibfnamefont {E.~J.}\ \bibnamefont
  {Kautz}}, \bibinfo {author} {\bibfnamefont {A.~R.}\ \bibnamefont {Hagen}},
  \bibinfo {author} {\bibfnamefont {J.~M.}\ \bibnamefont {Johns}}, \ and\
  \bibinfo {author} {\bibfnamefont {D.~E.}\ \bibnamefont {Burkes}},\
  }\href@noop {} {\bibfield  {journal} {\bibinfo  {journal} {Computational
  Materials Science}\ }\textbf {\bibinfo {volume} {161}},\ \bibinfo {pages}
  {107} (\bibinfo {year} {2019})}\BibitemShut {NoStop}%
\bibitem [{\citenamefont {R{\'e}da}\ \emph {et~al.}(2020)\citenamefont
  {R{\'e}da}, \citenamefont {Kaufmann},\ and\ \citenamefont
  {Delahaye-Duriez}}]{reda2020}%
  \BibitemOpen
  \bibfield  {author} {\bibinfo {author} {\bibfnamefont {C.}~\bibnamefont
  {R{\'e}da}}, \bibinfo {author} {\bibfnamefont {E.}~\bibnamefont {Kaufmann}},
  \ and\ \bibinfo {author} {\bibfnamefont {A.}~\bibnamefont
  {Delahaye-Duriez}},\ }\href@noop {} {\bibfield  {journal} {\bibinfo
  {journal} {Computational and Structural Biotechnology Journal}\ }\textbf
  {\bibinfo {volume} {18}},\ \bibinfo {pages} {241} (\bibinfo {year}
  {2020})}\BibitemShut {NoStop}%
\bibitem [{\citenamefont {Pal}\ \emph {et~al.}(2022)\citenamefont {Pal},
  \citenamefont {Park}, \citenamefont {Xia}, \citenamefont {Shen},\ and\
  \citenamefont {Wolverton}}]{pal2022}%
  \BibitemOpen
  \bibfield  {author} {\bibinfo {author} {\bibfnamefont {K.}~\bibnamefont
  {Pal}}, \bibinfo {author} {\bibfnamefont {C.~W.}\ \bibnamefont {Park}},
  \bibinfo {author} {\bibfnamefont {Y.}~\bibnamefont {Xia}}, \bibinfo {author}
  {\bibfnamefont {J.}~\bibnamefont {Shen}}, \ and\ \bibinfo {author}
  {\bibfnamefont {C.}~\bibnamefont {Wolverton}},\ }\href@noop {} {\bibfield
  {journal} {\bibinfo  {journal} {npj Computational Materials}\ }\textbf
  {\bibinfo {volume} {8}},\ \bibinfo {pages} {48} (\bibinfo {year}
  {2022})}\BibitemShut {NoStop}%
\bibitem [{\citenamefont {Hu}\ \emph {et~al.}(2020)\citenamefont {Hu},
  \citenamefont {Iwamoto}, \citenamefont {Feng}, \citenamefont {Ju},
  \citenamefont {Hu}, \citenamefont {Ohnishi}, \citenamefont {Nagai},
  \citenamefont {Hirakawa},\ and\ \citenamefont {Shiomi}}]{hu2020}%
  \BibitemOpen
  \bibfield  {author} {\bibinfo {author} {\bibfnamefont {R.}~\bibnamefont
  {Hu}}, \bibinfo {author} {\bibfnamefont {S.}~\bibnamefont {Iwamoto}},
  \bibinfo {author} {\bibfnamefont {L.}~\bibnamefont {Feng}}, \bibinfo {author}
  {\bibfnamefont {S.}~\bibnamefont {Ju}}, \bibinfo {author} {\bibfnamefont
  {S.}~\bibnamefont {Hu}}, \bibinfo {author} {\bibfnamefont {M.}~\bibnamefont
  {Ohnishi}}, \bibinfo {author} {\bibfnamefont {N.}~\bibnamefont {Nagai}},
  \bibinfo {author} {\bibfnamefont {K.}~\bibnamefont {Hirakawa}}, \ and\
  \bibinfo {author} {\bibfnamefont {J.}~\bibnamefont {Shiomi}},\ }\href
  {\doibase 10.1103/PhysRevX.10.021050} {\bibfield  {journal} {\bibinfo
  {journal} {Phys. Rev. X}\ }\textbf {\bibinfo {volume} {10}},\ \bibinfo
  {pages} {021050} (\bibinfo {year} {2020})}\BibitemShut {NoStop}%
\bibitem [{\citenamefont {Rodriguez}\ \emph {et~al.}(2023)\citenamefont
  {Rodriguez}, \citenamefont {Lin}, \citenamefont {Shen}, \citenamefont {Yuan},
  \citenamefont {Al-Fahdi}, \citenamefont {Zhang}, \citenamefont {Zhang},\ and\
  \citenamefont {Hu}}]{rodriguez2023}%
  \BibitemOpen
  \bibfield  {author} {\bibinfo {author} {\bibfnamefont {A.}~\bibnamefont
  {Rodriguez}}, \bibinfo {author} {\bibfnamefont {C.}~\bibnamefont {Lin}},
  \bibinfo {author} {\bibfnamefont {C.}~\bibnamefont {Shen}}, \bibinfo {author}
  {\bibfnamefont {K.}~\bibnamefont {Yuan}}, \bibinfo {author} {\bibfnamefont
  {M.}~\bibnamefont {Al-Fahdi}}, \bibinfo {author} {\bibfnamefont
  {X.}~\bibnamefont {Zhang}}, \bibinfo {author} {\bibfnamefont
  {H.}~\bibnamefont {Zhang}}, \ and\ \bibinfo {author} {\bibfnamefont
  {M.}~\bibnamefont {Hu}},\ }\href@noop {} {\bibfield  {journal} {\bibinfo
  {journal} {Communications Materials}\ }\textbf {\bibinfo {volume} {4}},\
  \bibinfo {pages} {61} (\bibinfo {year} {2023})}\BibitemShut {NoStop}%
\bibitem [{\citenamefont {Visaria}\ and\ \citenamefont
  {Jain}(2020)}]{visaria2020}%
  \BibitemOpen
  \bibfield  {author} {\bibinfo {author} {\bibfnamefont {D.}~\bibnamefont
  {Visaria}}\ and\ \bibinfo {author} {\bibfnamefont {A.}~\bibnamefont {Jain}},\
  }\href@noop {} {\bibfield  {journal} {\bibinfo  {journal} {Applied Physics
  Letters}\ }\textbf {\bibinfo {volume} {117}} (\bibinfo {year}
  {2020})}\BibitemShut {NoStop}%
\bibitem [{\citenamefont {Zhu}\ \emph {et~al.}(2021)\citenamefont {Zhu},
  \citenamefont {He}, \citenamefont {Gong}, \citenamefont {Xie}, \citenamefont
  {Gorai}, \citenamefont {Nielsch},\ and\ \citenamefont {Grossman}}]{zhu2021}%
  \BibitemOpen
  \bibfield  {author} {\bibinfo {author} {\bibfnamefont {T.}~\bibnamefont
  {Zhu}}, \bibinfo {author} {\bibfnamefont {R.}~\bibnamefont {He}}, \bibinfo
  {author} {\bibfnamefont {S.}~\bibnamefont {Gong}}, \bibinfo {author}
  {\bibfnamefont {T.}~\bibnamefont {Xie}}, \bibinfo {author} {\bibfnamefont
  {P.}~\bibnamefont {Gorai}}, \bibinfo {author} {\bibfnamefont
  {K.}~\bibnamefont {Nielsch}}, \ and\ \bibinfo {author} {\bibfnamefont
  {J.~C.}\ \bibnamefont {Grossman}},\ }\href@noop {} {\bibfield  {journal}
  {\bibinfo  {journal} {Energy \& Environmental Science}\ }\textbf {\bibinfo
  {volume} {14}},\ \bibinfo {pages} {3559} (\bibinfo {year}
  {2021})}\BibitemShut {NoStop}%
\bibitem [{\citenamefont {Liu}\ \emph {et~al.}(2022{\natexlab{a}})\citenamefont
  {Liu}, \citenamefont {Jiang},\ and\ \citenamefont {Luo}}]{liu2022}%
  \BibitemOpen
  \bibfield  {author} {\bibinfo {author} {\bibfnamefont {Z.}~\bibnamefont
  {Liu}}, \bibinfo {author} {\bibfnamefont {M.}~\bibnamefont {Jiang}}, \ and\
  \bibinfo {author} {\bibfnamefont {T.}~\bibnamefont {Luo}},\ }\href@noop {}
  {\bibfield  {journal} {\bibinfo  {journal} {Materials Today Physics}\
  }\textbf {\bibinfo {volume} {28}},\ \bibinfo {pages} {100868} (\bibinfo
  {year} {2022}{\natexlab{a}})}\BibitemShut {NoStop}%
\bibitem [{\citenamefont {Reissland}(1973)}]{reissland1973}%
  \BibitemOpen
  \bibfield  {author} {\bibinfo {author} {\bibfnamefont {J.~A.}\ \bibnamefont
  {Reissland}},\ }\href@noop {} {\emph {\bibinfo {title} {The Physics of
  Phonons}}}\ (\bibinfo  {publisher} {John Wiley and Sons Ltd},\ \bibinfo
  {year} {1973})\BibitemShut {NoStop}%
\bibitem [{\citenamefont {Srivastava}(1990)}]{srivastava1990}%
  \BibitemOpen
  \bibfield  {author} {\bibinfo {author} {\bibfnamefont {G.~P.}\ \bibnamefont
  {Srivastava}},\ }\href@noop {} {\emph {\bibinfo {title} {The Physics of
  Phonons}}}\ (\bibinfo  {publisher} {Adam Hilger, Bristol},\ \bibinfo {year}
  {1990})\BibitemShut {NoStop}%
\bibitem [{\citenamefont {Baroni}\ \emph {et~al.}(2001)\citenamefont {Baroni},
  \citenamefont {{de Gironcoli}}, \citenamefont {{Dal Corso}},\ and\
  \citenamefont {Giannozzi}}]{baroni2001}%
  \BibitemOpen
  \bibfield  {author} {\bibinfo {author} {\bibfnamefont {S.}~\bibnamefont
  {Baroni}}, \bibinfo {author} {\bibfnamefont {S.}~\bibnamefont {{de
  Gironcoli}}}, \bibinfo {author} {\bibfnamefont {A.}~\bibnamefont {{Dal
  Corso}}}, \ and\ \bibinfo {author} {\bibfnamefont {P.}~\bibnamefont
  {Giannozzi}},\ }\href@noop {} {\bibfield  {journal} {\bibinfo  {journal}
  {Reviews of Modern Physics}\ }\textbf {\bibinfo {volume} {73}},\ \bibinfo
  {pages} {515} (\bibinfo {year} {2001})}\BibitemShut {NoStop}%
\bibitem [{\citenamefont {Giannozzi}\ \emph {et~al.}(2009)\citenamefont
  {Giannozzi}, \citenamefont {Baroni}, \citenamefont {Bonini}, \citenamefont
  {Calandra}, \citenamefont {Car}, \citenamefont {Cavazzoni}, \citenamefont
  {Ceresoli}, \citenamefont {Chiarotti}, \citenamefont {Cococcioni},
  \citenamefont {Dabo}, \citenamefont {Corso}, \citenamefont {de~Gironcoli},
  \citenamefont {Fabris}, \citenamefont {Fratesi}, \citenamefont {Gebauer},
  \citenamefont {Gerstmann}, \citenamefont {Gougoussis}, \citenamefont
  {Kokalj}, \citenamefont {Lazzeri}, \citenamefont {Martin-Samos},
  \citenamefont {Marzari}, \citenamefont {Mauri}, \citenamefont {Mazzarello},
  \citenamefont {Paolini}, \citenamefont {Pasquarello}, \citenamefont
  {Paulatto}, \citenamefont {Sbraccia}, \citenamefont {Scandolo}, \citenamefont
  {Sclauzero}, \citenamefont {Seitsonen}, \citenamefont {Smogunov},
  \citenamefont {Umari},\ and\ \citenamefont {Wentzcovitch}}]{giannozzi2009}%
  \BibitemOpen
  \bibfield  {author} {\bibinfo {author} {\bibfnamefont {P.}~\bibnamefont
  {Giannozzi}}, \bibinfo {author} {\bibfnamefont {S.}~\bibnamefont {Baroni}},
  \bibinfo {author} {\bibfnamefont {N.}~\bibnamefont {Bonini}}, \bibinfo
  {author} {\bibfnamefont {M.}~\bibnamefont {Calandra}}, \bibinfo {author}
  {\bibfnamefont {R.}~\bibnamefont {Car}}, \bibinfo {author} {\bibfnamefont
  {C.}~\bibnamefont {Cavazzoni}}, \bibinfo {author} {\bibfnamefont
  {D.}~\bibnamefont {Ceresoli}}, \bibinfo {author} {\bibfnamefont {G.~L.}\
  \bibnamefont {Chiarotti}}, \bibinfo {author} {\bibfnamefont {M.}~\bibnamefont
  {Cococcioni}}, \bibinfo {author} {\bibfnamefont {I.}~\bibnamefont {Dabo}},
  \bibinfo {author} {\bibfnamefont {A.~D.}\ \bibnamefont {Corso}}, \bibinfo
  {author} {\bibfnamefont {S.}~\bibnamefont {de~Gironcoli}}, \bibinfo {author}
  {\bibfnamefont {S.}~\bibnamefont {Fabris}}, \bibinfo {author} {\bibfnamefont
  {G.}~\bibnamefont {Fratesi}}, \bibinfo {author} {\bibfnamefont
  {R.}~\bibnamefont {Gebauer}}, \bibinfo {author} {\bibfnamefont
  {U.}~\bibnamefont {Gerstmann}}, \bibinfo {author} {\bibfnamefont
  {C.}~\bibnamefont {Gougoussis}}, \bibinfo {author} {\bibfnamefont
  {A.}~\bibnamefont {Kokalj}}, \bibinfo {author} {\bibfnamefont
  {M.}~\bibnamefont {Lazzeri}}, \bibinfo {author} {\bibfnamefont
  {L.}~\bibnamefont {Martin-Samos}}, \bibinfo {author} {\bibfnamefont
  {N.}~\bibnamefont {Marzari}}, \bibinfo {author} {\bibfnamefont
  {F.}~\bibnamefont {Mauri}}, \bibinfo {author} {\bibfnamefont
  {R.}~\bibnamefont {Mazzarello}}, \bibinfo {author} {\bibfnamefont
  {S.}~\bibnamefont {Paolini}}, \bibinfo {author} {\bibfnamefont
  {A.}~\bibnamefont {Pasquarello}}, \bibinfo {author} {\bibfnamefont
  {L.}~\bibnamefont {Paulatto}}, \bibinfo {author} {\bibfnamefont
  {C.}~\bibnamefont {Sbraccia}}, \bibinfo {author} {\bibfnamefont
  {S.}~\bibnamefont {Scandolo}}, \bibinfo {author} {\bibfnamefont
  {G.}~\bibnamefont {Sclauzero}}, \bibinfo {author} {\bibfnamefont {A.~P.}\
  \bibnamefont {Seitsonen}}, \bibinfo {author} {\bibfnamefont {A.}~\bibnamefont
  {Smogunov}}, \bibinfo {author} {\bibfnamefont {P.}~\bibnamefont {Umari}}, \
  and\ \bibinfo {author} {\bibfnamefont {R.~M.}\ \bibnamefont {Wentzcovitch}},\
  }\href {http://stacks.iop.org/0953-8984/21/i=39/a=395502} {\bibfield
  {journal} {\bibinfo  {journal} {J Phys-Condens Mat}\ }\textbf {\bibinfo
  {volume} {21}},\ \bibinfo {pages} {395502} (\bibinfo {year}
  {2009})}\BibitemShut {NoStop}%
\bibitem [{\citenamefont {Schlipf}\ and\ \citenamefont
  {Gygi}(2015)}]{schlipf2015}%
  \BibitemOpen
  \bibfield  {author} {\bibinfo {author} {\bibfnamefont {M.}~\bibnamefont
  {Schlipf}}\ and\ \bibinfo {author} {\bibfnamefont {F.}~\bibnamefont {Gygi}},\
  }\href@noop {} {\bibfield  {journal} {\bibinfo  {journal} {Computer Physics
  Communications}\ }\textbf {\bibinfo {volume} {196}},\ \bibinfo {pages} {36}
  (\bibinfo {year} {2015})}\BibitemShut {NoStop}%
\bibitem [{\citenamefont {Jain}\ \emph {et~al.}(2013)\citenamefont {Jain},
  \citenamefont {Ong}, \citenamefont {Hautier}, \citenamefont {Chen},
  \citenamefont {Richards}, \citenamefont {Dacek}, \citenamefont {Cholia},
  \citenamefont {Gunter}, \citenamefont {Skinner}, \citenamefont {Ceder},\ and\
  \citenamefont {Persson}}]{materialsProject}%
  \BibitemOpen
  \bibfield  {author} {\bibinfo {author} {\bibfnamefont {A.}~\bibnamefont
  {Jain}}, \bibinfo {author} {\bibfnamefont {S.~P.}\ \bibnamefont {Ong}},
  \bibinfo {author} {\bibfnamefont {G.}~\bibnamefont {Hautier}}, \bibinfo
  {author} {\bibfnamefont {W.}~\bibnamefont {Chen}}, \bibinfo {author}
  {\bibfnamefont {W.~D.}\ \bibnamefont {Richards}}, \bibinfo {author}
  {\bibfnamefont {S.}~\bibnamefont {Dacek}}, \bibinfo {author} {\bibfnamefont
  {S.}~\bibnamefont {Cholia}}, \bibinfo {author} {\bibfnamefont
  {D.}~\bibnamefont {Gunter}}, \bibinfo {author} {\bibfnamefont
  {D.}~\bibnamefont {Skinner}}, \bibinfo {author} {\bibfnamefont
  {G.}~\bibnamefont {Ceder}}, \ and\ \bibinfo {author} {\bibfnamefont {K.~A.}\
  \bibnamefont {Persson}},\ }\href {\doibase 10.1063/1.4812323} {\bibfield
  {journal} {\bibinfo  {journal} {APL Materials}\ }\textbf {\bibinfo {volume}
  {1}},\ \bibinfo {pages} {011002} (\bibinfo {year} {2013})},\ \Eprint
  {http://arxiv.org/abs/https://pubs.aip.org/aip/apm/article-pdf/doi/10.1063/1.4812323/13163869/011002\_1\_online.pdf}
  {https://pubs.aip.org/aip/apm/article-pdf/doi/10.1063/1.4812323/13163869/011002\_1\_online.pdf}
  \BibitemShut {NoStop}%
\bibitem [{\citenamefont {Kingma}\ and\ \citenamefont {Ba}(2014)}]{kingma2014}%
  \BibitemOpen
  \bibfield  {author} {\bibinfo {author} {\bibfnamefont {D.~P.}\ \bibnamefont
  {Kingma}}\ and\ \bibinfo {author} {\bibfnamefont {J.}~\bibnamefont {Ba}},\
  }\href@noop {} {\bibfield  {journal} {\bibinfo  {journal} {arXiv preprint
  arXiv:1412.6980}\ } (\bibinfo {year} {2014})}\BibitemShut {NoStop}%
\bibitem [{\citenamefont {Xie}\ and\ \citenamefont {Grossman}(2018)}]{xie2017}%
  \BibitemOpen
  \bibfield  {author} {\bibinfo {author} {\bibfnamefont {T.}~\bibnamefont
  {Xie}}\ and\ \bibinfo {author} {\bibfnamefont {J.~C.}\ \bibnamefont
  {Grossman}},\ }\href {\doibase 10.1103/PhysRevLett.120.145301} {\bibfield
  {journal} {\bibinfo  {journal} {Phys. Rev. Lett.}\ }\textbf {\bibinfo
  {volume} {120}},\ \bibinfo {pages} {145301} (\bibinfo {year}
  {2018})}\BibitemShut {NoStop}%
\bibitem [{\citenamefont {Chen}\ \emph {et~al.}(2019)\citenamefont {Chen},
  \citenamefont {Ye}, \citenamefont {Zuo}, \citenamefont {Zheng},\ and\
  \citenamefont {Ong}}]{chen2019}%
  \BibitemOpen
  \bibfield  {author} {\bibinfo {author} {\bibfnamefont {C.}~\bibnamefont
  {Chen}}, \bibinfo {author} {\bibfnamefont {W.}~\bibnamefont {Ye}}, \bibinfo
  {author} {\bibfnamefont {Y.}~\bibnamefont {Zuo}}, \bibinfo {author}
  {\bibfnamefont {C.}~\bibnamefont {Zheng}}, \ and\ \bibinfo {author}
  {\bibfnamefont {S.~P.}\ \bibnamefont {Ong}},\ }\href@noop {} {\bibfield
  {journal} {\bibinfo  {journal} {Chemistry of Materials}\ }\textbf {\bibinfo
  {volume} {31}},\ \bibinfo {pages} {3564} (\bibinfo {year}
  {2019})}\BibitemShut {NoStop}%
\bibitem [{\citenamefont {Paszke}\ \emph {et~al.}(2019)\citenamefont {Paszke},
  \citenamefont {Gross}, \citenamefont {Massa}, \citenamefont {Lerer},
  \citenamefont {Bradbury}, \citenamefont {Chanan}, \citenamefont {Killeen},
  \citenamefont {Lin}, \citenamefont {Gimelshein}, \citenamefont {Antiga},
  \citenamefont {Desmaison}, \citenamefont {Kopf}, \citenamefont {Yang},
  \citenamefont {DeVito}, \citenamefont {Raison}, \citenamefont {Tejani},
  \citenamefont {Chilamkurthy}, \citenamefont {Steiner}, \citenamefont {Fang},
  \citenamefont {Bai},\ and\ \citenamefont {Chintala}}]{pytorch}%
  \BibitemOpen
  \bibfield  {author} {\bibinfo {author} {\bibfnamefont {A.}~\bibnamefont
  {Paszke}}, \bibinfo {author} {\bibfnamefont {S.}~\bibnamefont {Gross}},
  \bibinfo {author} {\bibfnamefont {F.}~\bibnamefont {Massa}}, \bibinfo
  {author} {\bibfnamefont {A.}~\bibnamefont {Lerer}}, \bibinfo {author}
  {\bibfnamefont {J.}~\bibnamefont {Bradbury}}, \bibinfo {author}
  {\bibfnamefont {G.}~\bibnamefont {Chanan}}, \bibinfo {author} {\bibfnamefont
  {T.}~\bibnamefont {Killeen}}, \bibinfo {author} {\bibfnamefont
  {Z.}~\bibnamefont {Lin}}, \bibinfo {author} {\bibfnamefont {N.}~\bibnamefont
  {Gimelshein}}, \bibinfo {author} {\bibfnamefont {L.}~\bibnamefont {Antiga}},
  \bibinfo {author} {\bibfnamefont {A.}~\bibnamefont {Desmaison}}, \bibinfo
  {author} {\bibfnamefont {A.}~\bibnamefont {Kopf}}, \bibinfo {author}
  {\bibfnamefont {E.}~\bibnamefont {Yang}}, \bibinfo {author} {\bibfnamefont
  {Z.}~\bibnamefont {DeVito}}, \bibinfo {author} {\bibfnamefont
  {M.}~\bibnamefont {Raison}}, \bibinfo {author} {\bibfnamefont
  {A.}~\bibnamefont {Tejani}}, \bibinfo {author} {\bibfnamefont
  {S.}~\bibnamefont {Chilamkurthy}}, \bibinfo {author} {\bibfnamefont
  {B.}~\bibnamefont {Steiner}}, \bibinfo {author} {\bibfnamefont
  {L.}~\bibnamefont {Fang}}, \bibinfo {author} {\bibfnamefont {J.}~\bibnamefont
  {Bai}}, \ and\ \bibinfo {author} {\bibfnamefont {S.}~\bibnamefont
  {Chintala}},\ }in\ \href
  {http://papers.neurips.cc/paper/9015-pytorch-an-imperative-style-high-performance-deep-learning-library.pdf}
  {\emph {\bibinfo {booktitle} {Advances in Neural Information Processing
  Systems 32}}}\ (\bibinfo  {publisher} {Curran Associates, Inc.},\ \bibinfo
  {year} {2019})\ pp.\ \bibinfo {pages} {8024--8035}\BibitemShut {NoStop}%
\bibitem [{\citenamefont {Pedregosa}\ \emph {et~al.}(2011)\citenamefont
  {Pedregosa}, \citenamefont {Varoquaux}, \citenamefont {Gramfort},
  \citenamefont {Michel}, \citenamefont {Thirion}, \citenamefont {Grisel},
  \citenamefont {Blondel}, \citenamefont {Prettenhofer}, \citenamefont {Weiss},
  \citenamefont {Dubourg}, \citenamefont {Vanderplas}, \citenamefont {Passos},
  \citenamefont {Cournapeau}, \citenamefont {Brucher}, \citenamefont {Perrot},\
  and\ \citenamefont {Duchesnay}}]{sklearn}%
  \BibitemOpen
  \bibfield  {author} {\bibinfo {author} {\bibfnamefont {F.}~\bibnamefont
  {Pedregosa}}, \bibinfo {author} {\bibfnamefont {G.}~\bibnamefont
  {Varoquaux}}, \bibinfo {author} {\bibfnamefont {A.}~\bibnamefont {Gramfort}},
  \bibinfo {author} {\bibfnamefont {V.}~\bibnamefont {Michel}}, \bibinfo
  {author} {\bibfnamefont {B.}~\bibnamefont {Thirion}}, \bibinfo {author}
  {\bibfnamefont {O.}~\bibnamefont {Grisel}}, \bibinfo {author} {\bibfnamefont
  {M.}~\bibnamefont {Blondel}}, \bibinfo {author} {\bibfnamefont
  {P.}~\bibnamefont {Prettenhofer}}, \bibinfo {author} {\bibfnamefont
  {R.}~\bibnamefont {Weiss}}, \bibinfo {author} {\bibfnamefont
  {V.}~\bibnamefont {Dubourg}}, \bibinfo {author} {\bibfnamefont
  {J.}~\bibnamefont {Vanderplas}}, \bibinfo {author} {\bibfnamefont
  {A.}~\bibnamefont {Passos}}, \bibinfo {author} {\bibfnamefont
  {D.}~\bibnamefont {Cournapeau}}, \bibinfo {author} {\bibfnamefont
  {M.}~\bibnamefont {Brucher}}, \bibinfo {author} {\bibfnamefont
  {M.}~\bibnamefont {Perrot}}, \ and\ \bibinfo {author} {\bibfnamefont
  {E.}~\bibnamefont {Duchesnay}},\ }\href@noop {} {\bibfield  {journal}
  {\bibinfo  {journal} {Journal of Machine Learning Research}\ }\textbf
  {\bibinfo {volume} {12}},\ \bibinfo {pages} {2825} (\bibinfo {year}
  {2011})}\BibitemShut {NoStop}%
\bibitem [{\citenamefont {Rasmussen}\ and\ \citenamefont
  {Williams}(2005)}]{GPR0}%
  \BibitemOpen
  \bibfield  {author} {\bibinfo {author} {\bibfnamefont {C.~E.}\ \bibnamefont
  {Rasmussen}}\ and\ \bibinfo {author} {\bibfnamefont {C.~K.~I.}\ \bibnamefont
  {Williams}},\ }\href@noop {} {\emph {\bibinfo {title} {Gaussian processes for
  machine learning}}},\ Adaptive Computation and Machine Learning series\
  (\bibinfo  {publisher} {MIT Press},\ \bibinfo {address} {London, England},\
  \bibinfo {year} {2005})\BibitemShut {NoStop}%
\bibitem [{\citenamefont {Schulz}\ \emph {et~al.}(2018)\citenamefont {Schulz},
  \citenamefont {Speekenbrink},\ and\ \citenamefont {Krause}}]{RBFandconst}%
  \BibitemOpen
  \bibfield  {author} {\bibinfo {author} {\bibfnamefont {E.}~\bibnamefont
  {Schulz}}, \bibinfo {author} {\bibfnamefont {M.}~\bibnamefont
  {Speekenbrink}}, \ and\ \bibinfo {author} {\bibfnamefont {A.}~\bibnamefont
  {Krause}},\ }\href@noop {} {\bibfield  {journal} {\bibinfo  {journal} {J.
  Math. Psychol.}\ }\textbf {\bibinfo {volume} {85}},\ \bibinfo {pages} {1}
  (\bibinfo {year} {2018})}\BibitemShut {NoStop}%
\bibitem [{\citenamefont {Miranda-Valdez}\ \emph {et~al.}(2022)\citenamefont
  {Miranda-Valdez}, \citenamefont {Viitanen}, \citenamefont {Mac~Intyre},
  \citenamefont {Puisto}, \citenamefont {Koivisto},\ and\ \citenamefont
  {Alava}}]{White1}%
  \BibitemOpen
  \bibfield  {author} {\bibinfo {author} {\bibfnamefont {I.~Y.}\ \bibnamefont
  {Miranda-Valdez}}, \bibinfo {author} {\bibfnamefont {L.}~\bibnamefont
  {Viitanen}}, \bibinfo {author} {\bibfnamefont {J.}~\bibnamefont
  {Mac~Intyre}}, \bibinfo {author} {\bibfnamefont {A.}~\bibnamefont {Puisto}},
  \bibinfo {author} {\bibfnamefont {J.}~\bibnamefont {Koivisto}}, \ and\
  \bibinfo {author} {\bibfnamefont {M.}~\bibnamefont {Alava}},\ }\href@noop {}
  {\bibfield  {journal} {\bibinfo  {journal} {Carbohydr. Polym.}\ }\textbf
  {\bibinfo {volume} {298}},\ \bibinfo {pages} {119921} (\bibinfo {year}
  {2022})}\BibitemShut {NoStop}%
\bibitem [{\citenamefont {Wehbe}\ \emph {et~al.}(2017)\citenamefont {Wehbe},
  \citenamefont {Hildebrandt},\ and\ \citenamefont {Kirchner}}]{White2}%
  \BibitemOpen
  \bibfield  {author} {\bibinfo {author} {\bibfnamefont {B.}~\bibnamefont
  {Wehbe}}, \bibinfo {author} {\bibfnamefont {M.}~\bibnamefont {Hildebrandt}},
  \ and\ \bibinfo {author} {\bibfnamefont {F.}~\bibnamefont {Kirchner}},\ }in\
  \href@noop {} {\emph {\bibinfo {booktitle} {2017 {IEEE} International
  Conference on Robotics and Automation ({ICRA})}}}\ (\bibinfo  {publisher}
  {IEEE},\ \bibinfo {year} {2017})\BibitemShut {NoStop}%
\bibitem [{\citenamefont {Ward}\ \emph {et~al.}(2018)\citenamefont {Ward},
  \citenamefont {Dunn}, \citenamefont {Faghaninia}, \citenamefont {Zimmermann},
  \citenamefont {Bajaj}, \citenamefont {Wang}, \citenamefont {Montoya},
  \citenamefont {Chen}, \citenamefont {Bystrom}, \citenamefont {Dylla},
  \citenamefont {Chard}, \citenamefont {Asta}, \citenamefont {Persson},
  \citenamefont {Snyder}, \citenamefont {Foster},\ and\ \citenamefont
  {Jain}}]{matminer}%
  \BibitemOpen
  \bibfield  {author} {\bibinfo {author} {\bibfnamefont {L.}~\bibnamefont
  {Ward}}, \bibinfo {author} {\bibfnamefont {A.}~\bibnamefont {Dunn}}, \bibinfo
  {author} {\bibfnamefont {A.}~\bibnamefont {Faghaninia}}, \bibinfo {author}
  {\bibfnamefont {N.~E.~R.}\ \bibnamefont {Zimmermann}}, \bibinfo {author}
  {\bibfnamefont {S.}~\bibnamefont {Bajaj}}, \bibinfo {author} {\bibfnamefont
  {Q.}~\bibnamefont {Wang}}, \bibinfo {author} {\bibfnamefont {J.~H.}\
  \bibnamefont {Montoya}}, \bibinfo {author} {\bibfnamefont {J.}~\bibnamefont
  {Chen}}, \bibinfo {author} {\bibfnamefont {K.}~\bibnamefont {Bystrom}},
  \bibinfo {author} {\bibfnamefont {M.}~\bibnamefont {Dylla}}, \bibinfo
  {author} {\bibfnamefont {K.}~\bibnamefont {Chard}}, \bibinfo {author}
  {\bibfnamefont {M.}~\bibnamefont {Asta}}, \bibinfo {author} {\bibfnamefont
  {K.}~\bibnamefont {Persson}}, \bibinfo {author} {\bibfnamefont {G.~J.}\
  \bibnamefont {Snyder}}, \bibinfo {author} {\bibfnamefont {I.}~\bibnamefont
  {Foster}}, \ and\ \bibinfo {author} {\bibfnamefont {A.}~\bibnamefont
  {Jain}},\ }\href {\doibase 10.1016/j.commatsci.2018.06.008} {\bibfield
  {journal} {\bibinfo  {journal} {Computational Materials Science}\ }\textbf
  {\bibinfo {volume} {152}},\ \bibinfo {pages} {60} (\bibinfo {year}
  {2018})}\BibitemShut {NoStop}%
\bibitem [{\citenamefont {Srivastava}\ \emph {et~al.}(2014)\citenamefont
  {Srivastava}, \citenamefont {Hinton}, \citenamefont {Krizhevsky},
  \citenamefont {Sutskever},\ and\ \citenamefont
  {Salakhutdinov}}]{srivastava2014}%
  \BibitemOpen
  \bibfield  {author} {\bibinfo {author} {\bibfnamefont {N.}~\bibnamefont
  {Srivastava}}, \bibinfo {author} {\bibfnamefont {G.}~\bibnamefont {Hinton}},
  \bibinfo {author} {\bibfnamefont {A.}~\bibnamefont {Krizhevsky}}, \bibinfo
  {author} {\bibfnamefont {I.}~\bibnamefont {Sutskever}}, \ and\ \bibinfo
  {author} {\bibfnamefont {R.}~\bibnamefont {Salakhutdinov}},\ }\href
  {http://jmlr.org/papers/v15/srivastava14a.html} {\bibfield  {journal}
  {\bibinfo  {journal} {Journal of Machine Learning Research}\ }\textbf
  {\bibinfo {volume} {15}},\ \bibinfo {pages} {1929} (\bibinfo {year}
  {2014})}\BibitemShut {NoStop}%
\bibitem [{\citenamefont {{\"O}zden}\ \emph {et~al.}(2023)\citenamefont
  {{\"O}zden}, \citenamefont {Zu{\~n}iga-Puelles}, \citenamefont {Kortus},
  \citenamefont {Gumeniuk},\ and\ \citenamefont {Himcinschi}}]{lit1}%
  \BibitemOpen
  \bibfield  {author} {\bibinfo {author} {\bibfnamefont {A.}~\bibnamefont
  {{\"O}zden}}, \bibinfo {author} {\bibfnamefont {E.}~\bibnamefont
  {Zu{\~n}iga-Puelles}}, \bibinfo {author} {\bibfnamefont {J.}~\bibnamefont
  {Kortus}}, \bibinfo {author} {\bibfnamefont {R.}~\bibnamefont {Gumeniuk}}, \
  and\ \bibinfo {author} {\bibfnamefont {C.}~\bibnamefont {Himcinschi}},\
  }\href@noop {} {\bibfield  {journal} {\bibinfo  {journal} {J. Raman
  Spectrosc.}\ }\textbf {\bibinfo {volume} {54}},\ \bibinfo {pages} {84}
  (\bibinfo {year} {2023})}\BibitemShut {NoStop}%
\bibitem [{\citenamefont {Wagner-Reetz}\ \emph {et~al.}(2014)\citenamefont
  {Wagner-Reetz}, \citenamefont {Kasinathan}, \citenamefont {Schnelle},
  \citenamefont {Cardoso-Gil}, \citenamefont {Rosner}, \citenamefont {Grin},\
  and\ \citenamefont {Gille}}]{lit2}%
  \BibitemOpen
  \bibfield  {author} {\bibinfo {author} {\bibfnamefont {M.}~\bibnamefont
  {Wagner-Reetz}}, \bibinfo {author} {\bibfnamefont {D.}~\bibnamefont
  {Kasinathan}}, \bibinfo {author} {\bibfnamefont {W.}~\bibnamefont
  {Schnelle}}, \bibinfo {author} {\bibfnamefont {R.}~\bibnamefont
  {Cardoso-Gil}}, \bibinfo {author} {\bibfnamefont {H.}~\bibnamefont {Rosner}},
  \bibinfo {author} {\bibfnamefont {Y.}~\bibnamefont {Grin}}, \ and\ \bibinfo
  {author} {\bibfnamefont {P.}~\bibnamefont {Gille}},\ }\href@noop {}
  {\bibfield  {journal} {\bibinfo  {journal} {Phys. Rev. B Condens. Matter
  Mater. Phys.}\ }\textbf {\bibinfo {volume} {90}} (\bibinfo {year}
  {2014})}\BibitemShut {NoStop}%
\bibitem [{\citenamefont {Guo}\ \emph {et~al.}(2021)\citenamefont {Guo},
  \citenamefont {Weng}, \citenamefont {Jiang}, \citenamefont {Zhu},
  \citenamefont {Li}, \citenamefont {Yuan}, \citenamefont {Yang}, \citenamefont
  {Zhang}, \citenamefont {Luo}, \citenamefont {Grin},\ and\ \citenamefont
  {Zhao}}]{lit3}%
  \BibitemOpen
  \bibfield  {author} {\bibinfo {author} {\bibfnamefont {K.}~\bibnamefont
  {Guo}}, \bibinfo {author} {\bibfnamefont {T.}~\bibnamefont {Weng}}, \bibinfo
  {author} {\bibfnamefont {Y.}~\bibnamefont {Jiang}}, \bibinfo {author}
  {\bibfnamefont {Y.}~\bibnamefont {Zhu}}, \bibinfo {author} {\bibfnamefont
  {H.}~\bibnamefont {Li}}, \bibinfo {author} {\bibfnamefont {S.}~\bibnamefont
  {Yuan}}, \bibinfo {author} {\bibfnamefont {J.}~\bibnamefont {Yang}}, \bibinfo
  {author} {\bibfnamefont {J.}~\bibnamefont {Zhang}}, \bibinfo {author}
  {\bibfnamefont {J.}~\bibnamefont {Luo}}, \bibinfo {author} {\bibfnamefont
  {Y.}~\bibnamefont {Grin}}, \ and\ \bibinfo {author} {\bibfnamefont {J.-T.}\
  \bibnamefont {Zhao}},\ }\href@noop {} {\bibfield  {journal} {\bibinfo
  {journal} {Mater. Today Phys.}\ }\textbf {\bibinfo {volume} {21}},\ \bibinfo
  {pages} {100480} (\bibinfo {year} {2021})}\BibitemShut {NoStop}%
\bibitem [{\citenamefont {Liu}\ \emph {et~al.}(2022{\natexlab{b}})\citenamefont
  {Liu}, \citenamefont {Liu}, \citenamefont {Wang}, \citenamefont {Liu},
  \citenamefont {Yu}, \citenamefont {Liu}, \citenamefont {Su},\ and\
  \citenamefont {Xia}}]{lit4}%
  \BibitemOpen
  \bibfield  {author} {\bibinfo {author} {\bibfnamefont {Q.}~\bibnamefont
  {Liu}}, \bibinfo {author} {\bibfnamefont {K.-F.}\ \bibnamefont {Liu}},
  \bibinfo {author} {\bibfnamefont {Q.-Q.}\ \bibnamefont {Wang}}, \bibinfo
  {author} {\bibfnamefont {X.-C.}\ \bibnamefont {Liu}}, \bibinfo {author}
  {\bibfnamefont {F.}~\bibnamefont {Yu}}, \bibinfo {author} {\bibfnamefont
  {J.}~\bibnamefont {Liu}}, \bibinfo {author} {\bibfnamefont {Y.-Y.}\
  \bibnamefont {Su}}, \ and\ \bibinfo {author} {\bibfnamefont {S.-Q.}\
  \bibnamefont {Xia}},\ }\href@noop {} {\bibfield  {journal} {\bibinfo
  {journal} {Acta Mater.}\ }\textbf {\bibinfo {volume} {230}},\ \bibinfo
  {pages} {117853} (\bibinfo {year} {2022}{\natexlab{b}})}\BibitemShut
  {NoStop}%
\bibitem [{\citenamefont {Qiu}\ \emph {et~al.}(2015)\citenamefont {Qiu},
  \citenamefont {Wu}, \citenamefont {Ke}, \citenamefont {Yang},\ and\
  \citenamefont {Zhang}}]{lit5}%
  \BibitemOpen
  \bibfield  {author} {\bibinfo {author} {\bibfnamefont {W.}~\bibnamefont
  {Qiu}}, \bibinfo {author} {\bibfnamefont {L.}~\bibnamefont {Wu}}, \bibinfo
  {author} {\bibfnamefont {X.}~\bibnamefont {Ke}}, \bibinfo {author}
  {\bibfnamefont {J.}~\bibnamefont {Yang}}, \ and\ \bibinfo {author}
  {\bibfnamefont {W.}~\bibnamefont {Zhang}},\ }\href@noop {} {\bibfield
  {journal} {\bibinfo  {journal} {Sci. Rep.}\ }\textbf {\bibinfo {volume}
  {5}},\ \bibinfo {pages} {13643} (\bibinfo {year} {2015})}\BibitemShut
  {NoStop}%
\bibitem [{\citenamefont {Toberer}\ \emph {et~al.}(2010)\citenamefont
  {Toberer}, \citenamefont {Zevalkink}, \citenamefont {Crisosto},\ and\
  \citenamefont {Snyder}}]{lit6}%
  \BibitemOpen
  \bibfield  {author} {\bibinfo {author} {\bibfnamefont {E.~S.}\ \bibnamefont
  {Toberer}}, \bibinfo {author} {\bibfnamefont {A.}~\bibnamefont {Zevalkink}},
  \bibinfo {author} {\bibfnamefont {N.}~\bibnamefont {Crisosto}}, \ and\
  \bibinfo {author} {\bibfnamefont {G.~J.}\ \bibnamefont {Snyder}},\
  }\href@noop {} {\bibfield  {journal} {\bibinfo  {journal} {Adv. Funct.
  Mater.}\ }\textbf {\bibinfo {volume} {20}},\ \bibinfo {pages} {4375}
  (\bibinfo {year} {2010})}\BibitemShut {NoStop}%
\bibitem [{\citenamefont {Zeier}\ \emph {et~al.}(2012)\citenamefont {Zeier},
  \citenamefont {Zevalkink}, \citenamefont {Schechtel}, \citenamefont
  {Tremel},\ and\ \citenamefont {Snyder}}]{lit7}%
  \BibitemOpen
  \bibfield  {author} {\bibinfo {author} {\bibfnamefont {W.~G.}\ \bibnamefont
  {Zeier}}, \bibinfo {author} {\bibfnamefont {A.}~\bibnamefont {Zevalkink}},
  \bibinfo {author} {\bibfnamefont {E.}~\bibnamefont {Schechtel}}, \bibinfo
  {author} {\bibfnamefont {W.}~\bibnamefont {Tremel}}, \ and\ \bibinfo {author}
  {\bibfnamefont {G.~J.}\ \bibnamefont {Snyder}},\ }\href@noop {} {\bibfield
  {journal} {\bibinfo  {journal} {J. Mater. Chem.}\ }\textbf {\bibinfo {volume}
  {22}},\ \bibinfo {pages} {9826} (\bibinfo {year} {2012})}\BibitemShut
  {NoStop}%
\bibitem [{\citenamefont {Ding}\ \emph {et~al.}(2016)\citenamefont {Ding},
  \citenamefont {Carrete}, \citenamefont {Li}, \citenamefont {Gao},\ and\
  \citenamefont {Yao}}]{lit8}%
  \BibitemOpen
  \bibfield  {author} {\bibinfo {author} {\bibfnamefont {G.}~\bibnamefont
  {Ding}}, \bibinfo {author} {\bibfnamefont {J.}~\bibnamefont {Carrete}},
  \bibinfo {author} {\bibfnamefont {W.}~\bibnamefont {Li}}, \bibinfo {author}
  {\bibfnamefont {G.~Y.}\ \bibnamefont {Gao}}, \ and\ \bibinfo {author}
  {\bibfnamefont {K.}~\bibnamefont {Yao}},\ }\href@noop {} {\bibfield
  {journal} {\bibinfo  {journal} {Appl. Phys. Lett.}\ }\textbf {\bibinfo
  {volume} {108}},\ \bibinfo {pages} {233902} (\bibinfo {year}
  {2016})}\BibitemShut {NoStop}%
\bibitem [{\citenamefont {Ding}\ \emph {et~al.}(2018)\citenamefont {Ding},
  \citenamefont {He}, \citenamefont {Cheng}, \citenamefont {Wang},\ and\
  \citenamefont {Li}}]{lit9}%
  \BibitemOpen
  \bibfield  {author} {\bibinfo {author} {\bibfnamefont {G.}~\bibnamefont
  {Ding}}, \bibinfo {author} {\bibfnamefont {J.}~\bibnamefont {He}}, \bibinfo
  {author} {\bibfnamefont {Z.}~\bibnamefont {Cheng}}, \bibinfo {author}
  {\bibfnamefont {X.}~\bibnamefont {Wang}}, \ and\ \bibinfo {author}
  {\bibfnamefont {S.}~\bibnamefont {Li}},\ }\href@noop {} {\bibfield  {journal}
  {\bibinfo  {journal} {J. Mater. Chem. C Mater. Opt. Electron. Devices}\
  }\textbf {\bibinfo {volume} {6}},\ \bibinfo {pages} {13269} (\bibinfo {year}
  {2018})}\BibitemShut {NoStop}%
\bibitem [{\citenamefont {Yue}\ \emph {et~al.}(2022)\citenamefont {Yue},
  \citenamefont {Xu}, \citenamefont {Zhao}, \citenamefont {Meng},\ and\
  \citenamefont {Dai}}]{lit10}%
  \BibitemOpen
  \bibfield  {author} {\bibinfo {author} {\bibfnamefont {T.}~\bibnamefont
  {Yue}}, \bibinfo {author} {\bibfnamefont {B.}~\bibnamefont {Xu}}, \bibinfo
  {author} {\bibfnamefont {Y.}~\bibnamefont {Zhao}}, \bibinfo {author}
  {\bibfnamefont {S.}~\bibnamefont {Meng}}, \ and\ \bibinfo {author}
  {\bibfnamefont {Z.}~\bibnamefont {Dai}},\ }\href@noop {} {\bibfield
  {journal} {\bibinfo  {journal} {Phys. Chem. Chem. Phys.}\ }\textbf {\bibinfo
  {volume} {24}},\ \bibinfo {pages} {4666} (\bibinfo {year}
  {2022})}\BibitemShut {NoStop}%
\bibitem [{\citenamefont {Han}\ \emph {et~al.}(2023)\citenamefont {Han},
  \citenamefont {Li}, \citenamefont {Liu}, \citenamefont {Li},\ and\
  \citenamefont {Wang}}]{MLFF}%
  \BibitemOpen
  \bibfield  {author} {\bibinfo {author} {\bibfnamefont {T.}~\bibnamefont
  {Han}}, \bibinfo {author} {\bibfnamefont {J.}~\bibnamefont {Li}}, \bibinfo
  {author} {\bibfnamefont {L.}~\bibnamefont {Liu}}, \bibinfo {author}
  {\bibfnamefont {F.}~\bibnamefont {Li}}, \ and\ \bibinfo {author}
  {\bibfnamefont {L.-W.}\ \bibnamefont {Wang}},\ }\href@noop {} {\bibfield
  {journal} {\bibinfo  {journal} {New J. Phys.}\ }\textbf {\bibinfo {volume}
  {25}},\ \bibinfo {pages} {093007} (\bibinfo {year} {2023})}\BibitemShut
  {NoStop}%
\end{thebibliography}
\bibliographystyle{plain}

\end{document}